\documentclass[journal]{IEEEtran}  % Option "lettersize" causes an "Unused global option(s)" warning.
\IEEEoverridecommandlockouts
\usepackage{amsmath,amsfonts}
\usepackage{amssymb}
\usepackage{hhline}
\usepackage{algorithmic}
\usepackage{algorithm}
\usepackage{array}
\usepackage[dvipsnames]{xcolor}
\ifCLASSOPTIONcompsoc
    \usepackage[caption=false, font=normalsize, labelfont=sf, textfont=sf]{subfig}
\else
\usepackage[caption=false, font=footnotesize]{subfig}
\fi
\usepackage{textcomp}
\usepackage{stfloats}
\usepackage{url}
\usepackage{verbatim}
\usepackage{graphicx}
\usepackage{cite}

\newcommand{\BigO}[1]{\mathcal{O}\left(#1\right)}
\newcommand{\algrule}[1][.2pt]{\par\vskip.5\baselineskip\hrule height #1\par\vskip.5\baselineskip}
\newcolumntype{P}[1]{>{\centering\arraybackslash}p{#1}}

\begin{document}

\title{One Transform To Compute Them All: Efficient Fusion-Based Full-Reference Video Quality Assessment}

\author{Abhinau K. Venkataramanan, Cosmin Stejerean, Ioannis Katsavounidis, and Alan C. Bovik, ~\IEEEmembership{Fellow,~IEEE}
        % <-this % stops a space
\thanks{This research was sponsored by a grant from Meta Video Infrastructure, and by grant number 2019844 for the National Science Foundation AI Institute for Foundations of Machine Learning (IFML).}}% <-this % stops a space

% The paper headers
\markboth{Arxiv Preprint, 2023}%
{Venkataramanan \MakeLowercase{\textit{et al.}}: One Transform to Compute Them All}

\maketitle

\begin{abstract}
The Visual Multimethod Assessment Fusion (VMAF) algorithm has recently emerged as a state-of-the-art approach to video quality prediction, that now pervades the streaming and social media industry. However, since VMAF requires the evaluation of a heterogeneous set of quality models, it is computationally expensive. Given other advances in hardware-accelerated encoding, quality assessment is emerging as a significant bottleneck in video compression pipelines. Towards alleviating this burden, we propose a novel Fusion of Unified Quality Evaluators (FUNQUE) framework, by enabling computation sharing and by using a transform that is sensitive to visual perception to boost accuracy. Further, we expand the FUNQUE framework to define a collection of improved low-complexity fused-feature models that advance the state-of-the-art of video quality performance with respect to both accuracy, by 4.2\% to 5.3\%, and computational efficiency, by factors of 3.8 to 11 times!
\end{abstract}


\section{Introduction}
\label{sec:introduction}
The COVID-19 pandemic has accelerated a boom in online video consumption that has engulfed the internet in recent years. Despite returning to in-person work, education, and socializing in 2022, the total volume of internet video grew by 24\% over 2021, accounting for more than 65\% of the global internet volume. The top five streaming platforms alone accounted for nearly half of all video traffic \cite{ref:sandvine_2023}. Given the stakes of satisfying billions of viewers, modeling the perceptual quality of videos has emerged as a crucial area of research. Moreover, given the rapid rise of Augmented and Virtual Reality (AR/VR) platforms such as the Metaverse and the popularity of video-sharing social media platforms such as Instagram and TikTok, the amount of visual media being uploaded to and downloaded from the internet is expected to grow exponentially.

Objective models of perceptual quality are used to drive key decisions in the process of video delivery that affect the end-user's quality of experience. Perhaps the most important quality-related decision in the video delivery pipeline is the ``encoding recipe.'' The ``encoding recipe'' refers to the set of parameters used by video codecs like AVC \cite{ref:avc}, HEVC \cite{ref:vp9}, and AV1 \cite{ref:av1} to encode a video clip before transmission. 

Typical Adaptive Bitrate (ABR) streaming pipelines control two main parameters that define their encoding recipes - the bitrate, which is related to compression level, and the encoding resolution. The encoding resolution refers to the resolution to which videos are scaled before compression, and it is typically lower than the source resolution. Therefore, viewers often experience a combination of scaling and compression-based distortions such as blocking, blurring, and banding. When low-quality scalers such as bilinear interpolation are used, distortion due to aliasing may also be apparent.

Depending on the encoding complexity of a given source content, a ``bitrate ladder'' is constructed by analyzing the rate-distortion tradeoff offered by several combinations of bitrates and encoding resolutions. The Dynamic Optimizer (DO) \cite{ref:dyn_optimizer} is an example of an algorithm that yields a perceptually-guided ``Pareto-optimal'' bitrate ladder. In fact, bitrate ladders constructed using DO are convex in the bitrate-distortion domain. The quality of the bitrate ladder depends on how well the rate-distortion tradeoff is characterized, which in turn relies on using an accurate model of perceptual quality. 

Perceptual quality models are also used to benchmark video codecs and steer their improvement and development. The improvement in performance offered by one codec over another is measured in terms of the Bjontegaard-Delta Rate (BD-Rate), which is calculated with respect to perceptual quality models such as Structural Similarity (SSIM) \cite{ref:ssim} and Visual Multimethod Assessment Fusion (VMAF) \cite{ref:vmaf} \cite{ref:msu_comp}, and PSNR, which is a pixel-fidelity quality model. In short, the BD-Rate quantifies the bitrate savings offered by one codec or bitrate ladder over another when encoding quality is held constant.

Constructing bitrate ladders using algorithms such as DO may require hundreds of resizing/encoding operations and quality evaluations per scene. A state-of-the-art (SOTA) video quality model that is very widely used in this process is VMAF. VMAF is a fusion-based quality model that uses the outputs of ``smaller'' quality models (called ``atom'' quality models) as features to compute the final quality score. Therefore, while VMAF is an accurate predictor of human subjective judgments, it comes at a higher computational cost than models like SSIM and Multiscale SSIM (MS-SSIM) \cite{ref:ms_ssim}. Coupled with the recent development of custom application-specific hardware for encoding and video processing \cite{ref:meta_hw} \cite{ref:google_hw}, this has led to the quality assessment step emerging as one of the computational bottlenecks.

\begin{figure*}[t]
    \centering
    \includegraphics[width=0.7\linewidth]{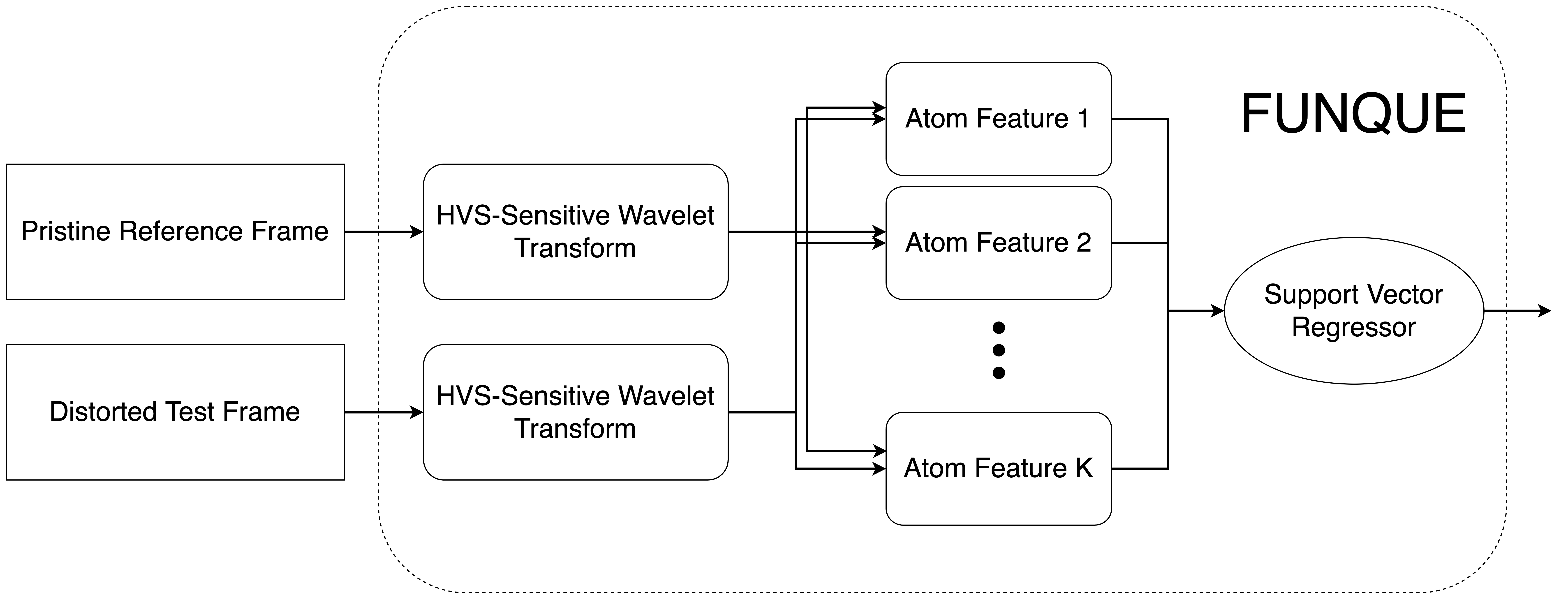}
    \caption{The FUNQUE framework for efficient fusion-based video quality modeling}
    \label{fig:funque_flowchart}
\end{figure*}

We propose novel fusion-based full-reference video quality models for streaming applications based on a new framework that we call Fusion of Unified Quality Evaluators (FUNQUE). In total, we make four novel contributions that culminate in the development of a suite of SOTA full-reference quality models that we call FUNQUE+:
\begin{enumerate}
    \item \textbf{The FUNQUE framework} - FUNQUE is a novel framework for full-reference quality modeling that we first introduced in \cite{ref:funque}. FUNQUE uses a perceptually tuned wavelet-domain transform that is shared by all the atom quality models that are to be fused. In this way, FUNQUE enables extensive computation sharing, leading to an overall low computational complexity while also achieving higher accuracy than other SOTA fusion-based models. A flowchart depicting FUNQUE is shown in Figure \ref{fig:funque_flowchart}, and a more detailed description is provided in Section \ref{sec:funque}.
    \item \textbf{Contrast sensitivity functions} - The shared wavelet transform is perceptually sensitized through the use of contrast sensitivity function (CSF) models of the visibility of visual artifacts. We deploy seven CSF models, some from the literature and others specifically designed for use in FUNQUE.
    \item \textbf{Atom quality models} - To facilitate the use of the shared transform, we design an extensive set of ``atom'' quality features that include both novel ``quality-aware'' features and others drawn from the literature. In particular, we demonstrate that the novel Multi-Scale Enhanced SSIM (MS-ESSIM) model designed here forms the backbone of FUNQUE+ models and is, by itself, a leading predictor of video quality. Part of the success of FUNQUE+ also derives from the redesign of several existing quality features as described in Section \ref{sec:features}, all computed using the shared wavelet transform coefficients.
    \item \textbf{Scalable feature selection} - While an extensive set of candidate features is desirable, we wanted the final fusion quality models to be compact and efficient. The use of traditional feature selection techniques such as exhaustive search or recursive feature elimination (RFE), also known as Greedy Feature Selection (GFS), is precluded by the large number of candidate features. Moreover, these techniques are not guaranteed to produce compact and efficient models. To overcome these limitations, we designed a Constrained Greedy Feature Selection (CGFS) method that uses ``feature buckets'' to scalably design compact fusion models.
\end{enumerate}

The remainder of this paper is organized as follows. Section \ref{sec:background} provides background regarding full-reference quality assessment, with a focus on recent fusion-based quality models, and Section \ref{sec:funque} provides a recap of the FUNQUE model. Section \ref{sec:csf} describes the various methods used to incorporate models of the CSF into the HVS-sensitive wavelet transform, and Section \ref{sec:features} describes the atom quality features extracted in the transform domain. Following this, Section \ref{sec:experiments} describes the experimental setup and the feature selection method used to develop our models, and Section \ref{sec:results} details the results of our experimental validation. Finally, Section \ref{sec:conclusion} concludes with a summary of our findings and possible directions for the future.
\section{Background}
\label{sec:background}
Objective models of image and video quality may be broadly categorized as Full-Reference (FR), No-Reference (NR), or Reduced-Reference (RR), depending on the availability of pristine reference contents. Since streaming services and video compression algorithms have direct access to pristine source content, we only consider the FR quality modeling problem here.

The Structural Similarity (SSIM) index \cite{ref:ssim} is a widely deployed perceptual objective FR video quality model used to control the quality of broadcast and streaming television content. The pervasive use of SSIM is due to the significant improvements it provides over the legacy PSNR model, which remains in use today, particularly in developing new video codec standards. The success of SSIM helped catalyze the development of more sophisticated FR quality models such as Multi-Scale SSIM (MS-SSIM) \cite{ref:ms_ssim}, the Feature Similarity index (FSIM) \cite{ref:fsim}, Visual Information Fidelity (VIF) \cite{ref:vif}, and the Detail Loss Metric (DLM) \cite{ref:dlm}, as well as reduced-reference models such as Spatio-Temporal Reduced Reference Entropic Differencing (ST-RRED) \cite{ref:strred}.

The aforementioned algorithms were generally designed based on models of perception and Natural Scene Statistics (NSS), and do not utilize data-driven machine-learning techniques. In a further advance, the Visual Multimethod Assessment Fusion (VMAF) \cite{ref:vmaf} model deployed fusion-based quality assessment. Fusion-based quality models employ a ``mixture-of-experts'' approach by using smaller learning-free quality models, called ``atom quality models,'' as features to train a machine-learning model, such as a Support Vector Regressor (SVR). Similar machine learning techniques, e.g., combining ``quality-aware'' features and models using SVRs was already common practice in the design of NR video quality algorithms, years prior to VMAF \cite{ref:biqi, ref:brisque, ref:diivine, ref:vbliinds}.

VMAF uses six features in total - scalar pixel-domain VIF computed at four Gaussian scales, DLM computed using a 4-level Db2 Discrete Wavelet Transform (DWT), and a ``Motion'' feature that characterizes the amount of temporal information present in the reference video. The computation of two 4-level multi-scale decompositions - the Gaussian pyramid in VIF and the Db2 DWT in DLM - are the primary reasons for VMAF's high computational complexity. Nevertheless, due to its improved accuracy as compared with the aforementioned models, VMAF has emerged as a SOTA FR perceptual video quality model in the streaming and compression space, alongside SSIM and MS-SSIM. Another major driver of VMAF's widespread adoption is the development of an efficient fixed-point implementation that significantly improves its computational complexity. \cite{ref:vmaf_fixed_point} While this technique is not explored here, fixed-point implementations may be, in principle, used to accelerate any of the models we discuss.

Attempts at improving VMAF have typically focused on adapting it as an optimization loss function in perceptual optimization problems or improving its accuracy. A key hurdle faced when attempting to use VMAF as a training objective for deep learning algorithms is its non-differentiability. To address this, differentiable approximations of VMAF, such as ProxIQA \cite{ref:prox_iqa} and ProxVQM \cite{ref:prox_vqm} have been developed, which adapt VMAF into a viable perceptual objective function to train deep image/video processing and compression algorithms. VMAF has also been used to predict Just Noticeable Differences (JND) \cite{ref:vmaf_jnd} for better design of ABR bitrate ladders. VMAF's atoms, and in particular VIF and DLM, were designed to capture both quality degradations and quality improvements and as such, VMAF carried over this feature, making it susceptible to reporting higher quality numbers after applying contrast enhancement and sharpening, as was demonstrated in \cite{ref:hacking_vmaf}. This prompted the development of the ``No Enhancement Gain'' VMAF model (VMAF-NEG) \cite{ref:ah_vmaf}, which is constrained to only measure quality degradations. 

On the other hand, attempts at improving VMAF's accuracy typically involve identifying better atom quality features. The Spatio-Temporal VMAF (ST-VMAF) and Ensemble VMAF (Ens-VMAF) \cite{ref:ensemble_vmaf} models introduce temporal quality-aware features derived from the Spatial Efficient Entropic Differencing (SpEED) quality models. In addition, Ens-VMAF uses an ensemble of two fusion models that use complementary features to boost accuracy. Enhanced VMAF (Enh-VMAF) \cite{ref:evmaf} further builds on this theme by incorporating temporal information using optical flow-based dynamic texture features (DTFs), ensemble modeling, and adding chroma-channel features to the feature pool.

VMAF has also found applications in domains other than standard HD video streaming, including chroma compression \cite{ref:vmaf_color}, 360 VR \cite{ref:vmaf_360vr}, high frame rate video \cite{ref:vmaf_hfr}, and medical videos \cite{ref:vmaf_medical}. Notably, all the aforementioned approaches to improving VMAF involve increasing the model size by adding more computationally complex features. Therefore, while these newer models may advance the SOTA in terms of accuracy, they exacerbate the computational cost problem faced by VMAF. Given the already considerable computational demands of VMAF, we will show that our FUNQUE and FUNQUE+ models are significantly less expensive than other fusion-based models, while also achieving higher accuracy.

Taking inspiration from the pooling strategy used by DLM, which computes quality over only the central 64\% area of an image, restricting the spatial region may lead to complexity improvements for any quality model. Furthermore, temporal subsampling \cite{ref:essim, ref:chipqa} has also emerged as a legitimate strategy for further reducing computational complexity, particularly when paired with scene-change-detection algorithms. While these methods are not discussed here, they may be applied to any of the quality models described here.

Deep-learning methods for quality modeling have gained popularity in recent years. Even before application to picture quality assessment, distances between VGG \cite{ref:vgg} features have been used as a perceptual loss to guide deep image generation tasks such as super-resolution and style-transfer \cite{ref:perceptual_loss, ref:superres_vgg}. The use of feature differences as a perceptual similarity metric was formalized by the Learned Perceptual Image Patch Similarity (LPIPS) \cite{ref:lpips} model, which uses a linear model to map differences between AlexNet \cite{ref:alexnet} features to similarity scores. Adversarially-robust versions of LPIPS such as E-LPIPS \cite{ref:e_lpips} and R-LPIPS \cite{ref:r_lpips} have also been developed using adversarial retraining. More recently, transformers have also been used to map differences between deep features to quality scores \cite{ref:lpips_transformer}, though these methods fall beyond the range of practical computational complexities of interest here.

The Deep Image Structure and Texture Similarity (DISTS) \cite{ref:dists} model adopts a different approach to comparing deep feature maps. The mean and standard deviation of VGG feature maps are calculated and compared using SSIM-like metrics, which function as perceptual distance measures. Finally, the Deep Wasserstein Distance (DeepWSD) model \cite{ref:deep_wsd} compares feature maps using statistical similarity metrics instead of pixel similarity/difference metrics. In particular, DeepWSD estimates the Wasserstein distance between the distributions of deep feature coefficients obtained from a pair of compared images.
\section{FUNQUE}
\label{sec:funque}
The foundation of the FUNQUE framework \cite{ref:funque} is a ``unified transform'' that is designed to be sensitive to properties of the Human Visual System (HVS). The role of this transform is two-fold. First, reusing the output of the unified transform for all of the atom quality models improves efficiency through computation sharing. Secondly, incorporating HVS-sensitivity into all the atom quality models improves accuracy against human judgments of perceptual video quality.

The latter is achieved by combining models of contrast sensitivity with the multi-scale frequency-selectivity of discrete wavelet transforms (DWTs). In particular, FUNQUE uses a 21-tap spatial filter based on the CSF model proposed in \cite{ref:ngan}, and a Haar wavelet transform to obtain the final transform coefficients. A detailed treatment of the CSF model used in FUNQUE, and the new models developed for FUNQUE+, is provided in Section \ref{sec:csf}.

The FUNQUE model designed in \cite{ref:funque} also uses the Self Adaptive Scale Transform (SAST) \cite{ref:sast}, which is a technique that involves rescaling reference and test input frames prior to quality estimation to account for viewing conditions. Similar to the use of the shared unified transform, SAST also serves the dual purpose of simultaneously improving efficiency and accuracy.

The transformed wavelet coefficients are then used to compute atom quality features for fusion. While FUNQUE is a general framework that may be used with any wavelet-domain features, the model proposed in \cite{ref:funque} used Enhanced SSIM (ESSIM) \cite{ref:essim}, VIF \cite{ref:vif}, DLM \cite{ref:dlm} and motion as the atom quality features. Through an ablation study, it was found that ESSIM in particular contributed over 50\% of the observed improvement in accuracy. Finally, these features were fused using a non-linear SVR model. A detailed description of the features used in FUNQUE, and the new features developed for FUNQUE+, is provided in Section \ref{sec:features}.

In summary, FUNQUE provides a framework for efficient, accurate fusion-based FR video quality models. The prototype model\footnote{The prototype FUNQUE model was presented at IEEE ICIP 2022.} presented in \cite{ref:funque} provides a proof-of-concept, but the development and analysis of the model left significant room for improvement, which is addressed here. The feature set that defines FUNQUE consists of pre-existing features that were primarily borrowed from VMAF. By contrast, here we develop a much wider range of novel features with an emphasis on efficiency and wavelet-domain compatibility while incorporating chromatic and spatio-temporal features.

Furthermore, the CSF model used in the original FUNQUE, which was used to demonstrate the impact of the unified transform, was directly borrowed from DLM \cite{ref:dlm}. Here, we conduct a much more extensive survey and analysis of efficient CSF models, and develop some of our own, towards developing a better Unified Transform model. Finally, we provide a significantly broader and deeper analysis of all the selected models, including extensive cross-database testing, statistical significance tests, computational complexity analysis, and monotonicity analysis. In this manner, we demonstrate significant advances over the prototype FUNQUE model \cite{ref:funque} and are able to show that FUNQUE+ significantly exceeds the performances of state-of-the-art FR video quality models.

\section{Contrast Sensitivity and the Unified Transform}
\label{sec:csf}
\subsection{The FUNQUE Unified Transform}

Broadly, the HVS-sensitive unified transform involves filtering the input frames using a model of human contrast sensitivity, followed by processing them with a suitable wavelet transform to conduct perceptually sensitized multi-scale analysis. We deploy a model of the CSF to capture information regarding the visibility of artifacts as a function of spatial frequency. In fact, we tested and compared several CSF models, as described in the following.

FUNQUE is a general modular framework that is based on a unified transform \cite{ref:funque} modulated by a simple model of contrast sensitivity, proposed by Ngan et al. \cite{ref:ngan}:
\begin{equation}
    \text{NganCSF}(f) = (0.31 + 0.69f) e^{-0.29f},
    \label{eq:ngan_csf}
\end{equation}
where \(f\) is expressed in units of cycles/degree.

Since applying Fourier transforms on HD frames is computationally expensive, the Ngan CSF model can be translated into a spatial (angle) domain filter by computing an analytical inverse Fourier Transform:
\begin{equation}
    \text{NganSpat}(\theta) = \frac{2\left(0.0656 - 23.6910\:\theta^2\right)}{\left(0.0841 + 39.4784\:\theta^2\right)^2}.
    \label{eq:ngan_spat}
\end{equation}

When viewing video content on a 1080p display of height \(H\), placed at a distance \(D\), the angle subtended by a pixel at the eye is
\begin{equation}
\Delta \theta \approx \tan(\Delta \theta) = \frac{H}{D \times 1080} \times \frac{180}{\pi}
\end{equation}
in degrees, which is the interval at which the above ``continuous-angle'' filter must be sampled. The corresponding discrete spatial domain filter is
\begin{equation}
    \text{NganSpat}[n] = \Delta \theta \times \text{NganSpat}(n \Delta \theta).
\end{equation}
Since FUNQUE was trained for TV viewing, a nominal value of \(D/H = 3.0\) was used, as in VMAF. Since the discrete spatial filter has infinite support, it was truncated to 21 taps, at which point, the value of the tail is less than 5\% of the peak value. 

\subsection{Towards Improved HVS Modeling}
A key drawback of FUNQUE's unified transform is the use of a 21-tap filter, which runs contrary to the goal of computational efficiency. The spatial filter form of the Ngan CSF, as used in the DLM algorithm \cite{ref:dlm}, assigns a multiplicative weight to each subband. The Li subband weight (LiSW) is computed as a function of the subband \(\theta \in \{H, V, D\}\), where \(H, V, D\) denote horizontal, vertical, and diagonal wavelet subbands, indexed by wavelet levels \(\lambda = 1, 2, \dots\)
\begin{equation}
    \text{LiSW}(\lambda, \theta) = \text{NganCSF}\left(f_{nom}(\lambda, \theta)\right),
    \label{eq:li_sw}
\end{equation}
where \(f_{nom}(\lambda, \theta)\) is the nominal frequency assigned to subband \(\theta\) at level \(\lambda\):
\begin{equation}
    f_{nom}(\lambda, \theta) = \frac{1080 \times \pi \times (D/H)}{2^{\lambda} \times 180 \times (0.15p(\theta) + 0.85)},
    \label{eq:f_nom}
\end{equation}
where \(p(\theta) = 1\) if \(\theta \in \{H, V\}\), and \(p(\theta) = -1\) if \(\theta = D\). This captures the oblique effect \cite{ref:oblique}, whereby the HVS is less sensitive to diagonal subbands than horizontal and vertical subbands.

In addition to the Ngan CSF model, we considered two other frequency-domain models. The second CSF model we considered was developed by Nadenau et al. \cite{ref:nadenau}, who used it to augment wavelet-domain color image compression. Since images are typically compressed in opponent color spaces, three models of CSF were developed in \cite{ref:nadenau}, corresponding to the ``luminance,'' ``red-green,'' and ``blue-yellow'' channels. All three CSFs are described in the frequency domain by the functional form
\begin{equation}
    \text{NadenauCSF}(f) = \left(1 + 255e^{-bf^{c}}\right)/256,
    \label{eq:nadenau_csf}
\end{equation}
where \(b\) and \(c\) vary across three channels. Note that unlike the NganCSF, which is bandpass, the Nadenau CSF has a lowpass response in all three channels. As with the NganCSF, we would like to apply the NadenauCSF as a spatial filter (NadenauSpat) to avoid computing Fourier Transforms. An analytical inverse proved difficult to compute due to the presence of \(f^{c}\) in the exponential function. Therefore, numerical inverses were calculated using the Fast Fourier Transform (FFT). The filters were truncated to yield 5\% error in the tail, leading to 5-tap filters for the Y and Cr channels and a 7-tap filter for the Cb channel. Note that all three NadenauSpat CSF filters are significantly more compact than the 21-tap NganSpat CSF filter.

Similar to the approach followed by Li et al. \cite{ref:dlm}, we also consider a subband-weighted version of the Nadenau CSF as a coarse yet efficient method of incorporating HVS-sensitivity. The Nadenau subband weights are computed as
\begin{equation}
    \text{NadenauSW}(\lambda, \theta) = \text{NadenauCSF}\left(f_{nom}(\lambda, \theta)\right),
    \label{eq:nadenau_sw}
\end{equation}
where \(f_{nom}\) is assigned as in (\ref{eq:f_nom}).

A third model of the CSF that we considered was formulated by Larson et al. \cite{ref:larson}, which is a modified version of the classic Mannos-Sakrison CSF model \cite{ref:mannos}. Larson's model is expressed as a function of radial spatial frequency \(f_r\) and orientation \(\phi\):
\begin{align}
    \text{Lars}&\text{onCSF}(f_r, \phi) \nonumber \\
    &= \begin{cases}
        (0.0499 + 0.5928f_\phi)e^{(0.228f_\phi)^{1.1}}, f_r \geq 4 \\
        0.981, \quad f_r < 4
    \end{cases},
\end{align}
where \(f_\phi = f_r/(0.15\cos(4\phi) + 0.85)\) is a corrected radial frequency that accounts for the oblique effect. Since the LarsonCSF does not yield a small spatial filter when transformed into the spatial domain, we only consider the subband-weighting method, with weights given by
\begin{equation}
    \text{LarsonSW}(f, \theta) = \text{LarsonCSF}(f_{nom}(\lambda, \theta), 0).
\end{equation}

Once again, the nominal frequencies were computed as in (\ref{eq:f_nom}), and the orientation \(\phi\) was set to zero since the oblique effect is already accounted for by \(f_{nom}\).

Note that all the aforementioned CSFs encode the visibility of various spatial frequencies. However, since we aim to use wavelet decompositions of the two input frames, we also considered models of wavelet subband visibility. In particular, we consider two models of wavelet subband CSFs, by Watson et al. \cite{ref:watson} and Hill et al. \cite{ref:hill}. Since both these wavelet CSF models are applied as subband-weighting mechanisms, we will refer to them as the Watson SW and Hill SW CSF models.

Therefore, in summary, we have experimented with seven methods of applying CSFs as part of defining unified transforms in the FUNQUE framework. Key details regarding the CSF methods used in our experiments are summarized in Table~\ref{tab:csfs}.

\begin{table}[t]
    \centering
    \caption{Summary of CSF Methods}
    \label{tab:csfs}
    \begin{tabular}{|c|c|c|c|}
    \hline
     Method & Base CSF Model & Domain & Color Aware?\\
    \hline
     NganSpat & Ngan \cite{ref:ngan} & Pixel & No \\
     LiSW & Ngan \cite{ref:ngan} & Wavelet & No \\
     NadenauSpat & Nadenau \cite{ref:nadenau} & Pixel & Yes \\
     NadenauSW & Nadenau \cite{ref:nadenau} & Wavelet & Yes \\
     LarsonSW & Larson \cite{ref:larson} & Wavelet & No \\
     WatsonSW & Watson \cite{ref:watson} & Wavelet & Yes \\
     HillSW & Hill \cite{ref:hill} & Wavelet & No\\
    \hline
    \end{tabular}
\end{table}

\subsection{Self Adaptive Scale Transform}
The Self-Adaptive Scale Transform (SAST) is a preprocessing method proposed in \cite{ref:sast} to account for viewing conditions, which are described by the viewing distance and the height of the display. When SAST is applied prior to quality assessment, both the reference and test video frames are downscaled by a factor of
\begin{equation}
    \alpha_\text{SAST} = \sqrt{4\tan\left(\theta_H/2\right)\tan\left(\theta_W/2\right) \cdot \frac{D}{H} \cdot \frac{D}{W}}
\end{equation}
where \(\theta_W\) and \(\theta_H\) denote the angular width and height of the field of view, \(W\) and \(H\) denote the physical width and height of the display, and \(D\) denotes the viewing distance.

Assuming a standard 16:9 display, and using typical values of \(\theta_H = 40^{\circ}\) and \(\theta_W=50^{\circ}\), then
\begin{equation}
    \alpha_\text{SAST} \approx \frac{D/H}{1.618}.
\end{equation}

In the following developments, we assume a \(D/H\) ratio of 3.0, which corresponds to typical TV viewing conditions. For simplicity, we round \(\alpha_\text{SAST}\) to the nearest integer, yielding a scale factor of two. As described in Section \ref{sec:results}, we have found that applying SAST improves the accuracy of the FUNQUE+ models. This observation mirrors the improvement in accuracy observed by both ESSIM and FUNQUE due to the application of SAST \cite{ref:funque, ref:essim}.

However, the use of SAST introduces a vulnerability in the FUNQUE+ model that may be exploited when it is used as a performance metric in applications such as perceptual optimization and codec evaluation. For example, since both reference and test frames are processed at half-resolution by FUNQUE+, a codec may achieve high (or even perfect) FUNQUE+ scores by also processing frames at half-resolution. However, this would clearly not lead to lossless coding since operating at half-resolution is inherently lossy. Therefore, we also report ``full-scale'' (FS) FUNQUE+ models that do not utilize SAST. Such models typically achieve lower accuracies but do not suffer the same vulnerability.
\section{Feature Extraction From The Unified Transform Domain}
\label{sec:features}

The FUNQUE framework for quality modeling is founded on the use of a shared transform from which all atom quality features are computed. Since the shared transform is based on a perceptually-sensitive wavelet transform, careful development of atom quality models is needed to enable maximum computation sharing. When adopting atom quality models from the literature, wavelet-domain models such as the Detail Loss Metric (DLM) \cite{ref:dlm} may be used off-the-shelf. However, spatial-domain models are not compatible with FUNQUE.

To adopt spatial domain models, we identify analogous quantities that may be computed in the wavelet domain. For example, horizontal and vertical subbands may be used as substitutes for gradients in gradient-based methods, and local means and variances within square windows may be computed from wavelet subbands using an approach similar to \cite{ref:funque}, which will also be explained in Section \ref{sec:ms_essim}. Such adaptations are crucial to the success of FUNQUE and FUNQUE+, and in this section, we describe the use of these techniques to develop our candidate feature set.

\subsection{Multi-Scale Enhanced Structural Similarity}
\label{sec:ms_essim}
As shown in \cite{ref:funque}, a key driver of the accuracy of the prototype model FUNQUE is the use of Enhanced Structural Similarity (ESSIM) \cite{ref:essim}. In its original form, ESSIM was computed in the spatial domain similar to SSIM. However, since the FUNQUE framework is predicated on computing all features from wavelet decompositions of {input frames from the reference and test videos, we have developed a method in \cite{ref:funque} to compute spatial-domain structural similarity (SSIM) directly from Haar wavelet coefficients. Local SSIM scores obtained in this manner are used to compute Enhanced SSIM \cite{ref:essim}, which uses the Coefficient of Variation (CoV) to conduct spatial aggregation and has been shown to outperform baseline SSIM. In this subsection, we review the wavelet-domain computation of ESSIM and propose a novel wavelet-domain Multi-Scale ESSIM (MS-ESSIM) quality model.

Consider a \(2\times2\) block of an input image \(x\). A 1-level Haar transform of this block may be expressed as
\begin{equation}
    \begin{bmatrix}
        X_{1,A} \\
        X_{1,H} \\
        X_{1,V} \\
        X_{1,D}
    \end{bmatrix} = \frac{1}{2}\begin{bmatrix}
        +1, +1, +1, +1 \\
        +1, -1, +1, -1 \\
        +1, +1, -1, -1 \\
        +1, -1, -1, +1 \\
    \end{bmatrix} \begin{bmatrix}
        x(i,j) \\
        x(i,j+1) \\
        x(i+1,j) \\
        x(i+1,j+1)
    \end{bmatrix}.
\end{equation}

Consider a pair of images \(x\) and \(y\) of size \(M \times N\) such that both \(M\) and \(N\) are divisible by \(2^L\). Assuming \(L\)-level Haar transforms of each, let \(X_{\lambda, \theta}\) and \(Y_{\lambda, \theta}\) denote the subband \(\theta \in \{A, H, V, D\}\), which corresponds to the approximation, horizontal, vertical, and diagonal subbands, at level \(\lambda \in \{1, \dots, L\}\). Since the transformation matrix is a scaled orthogonal matrix, inner products between \(2\times2\) blocks of the input images may be obtained from inner products between their corresponding wavelet coefficients with appropriate scaling. This argument may be extended further to include \(2^L \times 2^L\) blocks when using an \(L\)-level Haar transform.

Furthermore, local variances of images and covariances between images may be expressed as scaled inner products between corresponding blocks. Therefore, using the properties described above, local statistics from disjoint \(2^L \times 2^L\) blocks may be obtained from Haar wavelet coefficients as follows:
\begin{equation}
    \mu_{x, L}(i, j) = 2^{-L}X_{L, A}(i, j),
\end{equation}
\begin{equation}
    \mu_{y, L}(i, j) = 2^{-L}Y_{L, A}(i, j),
\end{equation}
\begin{equation}
    \sigma^2_{x, L}(i, j) = 2^{-2L}\sum_{k=1}^{L} \sum_{P^k_{ij}} \sum_{\{H, V, D\}} X_{k, \theta}^2(m, n),
\end{equation}
\begin{equation}
    \sigma^2_{y, L}(i, j) = 2^{-2L}\sum_{k=1}^{L} \sum_{P^k_{ij}} \sum_{\{H, V, D\}} Y_{k, \theta}^2(m, n),
\end{equation}
\begin{equation}
    \sigma_{xy, L}(i, j) = 2^{-2L}\sum_{k=1}^{L} \sum_{P^k_{ij}} \sum_{\{H, V, D\}} X_{k, \theta}(m, n) Y_{k, \theta}(m, n),
\end{equation}
where \(P^k_{ij} = \{(m, n) \mid i2^{L-k} \leq m < (i+1)2^{L-k}, j2^{L-k} \leq n < (j+1)2^{L-k}\}\) denotes disjoint \(2^{L-k}\times2^{L-k}\) blocks. These local statistics may be used to compute both SSIM and ESSIM in the wavelet domain, as in \cite{ref:essim}.

We extend this framework to compute MS-SSIM and MS-ESSIM by computing the aforementioned statistics at all levels \(\lambda \leq L\), as opposed to only computing them at level \(L\). Our multi-scale definition of wavelet-domain MS-(E)SSIM corresponds to computing spatial-domain (E)SSIM using windows that double in size between successive scales. This is analogous to the method used by traditional MS-SSIM, where images are downsampled by a factor of two between successive scales. Altering the multi-scale computation in this manner allows the computation of statistics at each scale in an iterative manner as shown below

\begin{align}
    \sigma_{x, \lambda+1}^2(i,j) &= 2^{-2(\lambda+1)}\sum_{k=1}^{\lambda+1} \sum_{P^k_{ij}} \sum_{\{H, V, D\}} X^2_{k, \theta}(m, n) \\
    &= 2^{-2}\sum\limits_{m=i}^{i+1}\sum\limits_{n=j}^{j+1}\sigma_{x, \lambda}(m,n)^2 + \Tilde{\sigma}^2_{x, \lambda+1}(i,j),
\end{align}
where
\begin{align}
    \Tilde{\sigma}^2_{x, \lambda+1}(i,j) = 2^{-2(\lambda+1)} \sum_{\{H, V, D\}} X^2_{\lambda+1, \theta}(i, j).
\end{align}

A similar iterative method is used to compute \(\sigma_{y, \lambda}^2(i,j)\) and \(\sigma_{xy, \lambda}(i,j)\) at each level from the previous levels. In this manner, we obtain local statistics at all scales using the same number of operations that computing single-scale (E)SSIM from an \(L\)-level wavelet decomposition would require.

Using the multi-scale local statistics computed in this manner, pooled luminance and contrast-structure similarity scores at each level may be obtained as
\begin{align}
    l_{\text{Pool}, \lambda} = \text{Pool}\left(\frac{2\mu_{x, \lambda}\mu_{y, \lambda} + K_1}{\mu_{x, \lambda}^2 + \mu_{y, \lambda}^2 + K_1}\right) \\
    cs_{\text{Pool}, \lambda} = \text{Pool}\left(\frac{2\sigma_{xy, \lambda} + K_2}{\sigma_{x, \lambda}^2 + \sigma_{y, \lambda}^2 + K_2}\right),
\end{align}
from which the level-\(L\) SSIM and ESSIM are computed by pooling the products of the luminance and contrast-similarity scores from level \(L\). Note that the Pool function is set to mean pooling when computing (MS-)SSIM, and to CoV pooling when computing (MS-)ESSIM. The final MS-SSIM and MS-ESSIM scores are obtained by combining scores across levels using the MS-SSIM exponents \(\alpha_\lambda\) \cite{ref:ms_ssim}:
\begin{align}
    \text{MS-SSIM}_L = \left(\text{SSIM}_{L}\right)^{\alpha_L} \prod_{\lambda = 1}^{L-1} \left(cs_{\text{Mean}, \lambda}\right)^{\alpha_\lambda} \\
    \text{MS-ESSIM}_L = \left(\text{ESSIM}_{L}\right)^{\alpha_L} \prod_{\lambda = 1}^{L-1} \left(cs_{\text{CoV}, \lambda}\right)^{\alpha_\lambda}.
\end{align}

\subsection{Information-Theoretic Features}
\label{sec:info}
As in FUNQUE, we use multi-scale Visual Information Fidelity (VIF) \cite{ref:vif} features. The VIF features, which are derived under a scalar Gaussian Scale Mixture (GSM) model, are computed from wavelet subbands at each level. In short, local means (\(\mu_{x, \lambda, \theta}, \mu_{y, \lambda, \theta}\)), variances (\(\sigma^2_{x, \lambda, \theta}, \sigma^2_{y, \lambda, \theta}\)), and covariances (\(\sigma_{xy, \lambda, \theta}\)) from subbands are computed using highly efficient integral images \cite{ref:essim}. Note that this notation extends the denotation of local statistics in Section \ref{sec:ms_essim}, where \(\lambda = 1 \dots L\) denotes the wavelet level, and \(\theta \in \{A, H, V, D\}\) denotes the wavelet subband. For example, \(\mu_{x, 1, A}(i,j)\) denotes local means computed from the approximation subband in the first level of the wavelet decomposition of the reference image.

As described in \cite{ref:vif}, VIF assumes that distortions arise from a channel that is modeled as
\begin{equation}
    Y_{\lambda, \theta}(i,j) = g_{\lambda, \theta}(i,j) X_{\lambda, \theta}(i, j) + N_{\lambda, \theta}(i, j),
\end{equation}
where \(g_{\lambda, \theta}(i,j)\) is the gain of the distortion channel and \(N_{\lambda, \theta}(i, j) \sim \mathcal{N}(0, \sigma^2_v)\) is additive white Gaussian noise.

As described in \cite{ref:vif}, local statistics are used to estimate channel parameters as:
\begin{equation}
    g_{\lambda, \theta}(i,j) = \sigma_{xy, \lambda, \theta}(i,j) / \sigma^2_{x, \lambda, \theta}(i,j)
\end{equation}
\begin{equation}
    \sigma^2_{v, \lambda, \theta}(i,j) = \sigma^2_{y, \lambda, \theta}(i,j) - g_{\lambda, \theta}(i,j) \sigma_{xy, \lambda, \theta}(i,j)
\end{equation}

The estimated channel parameters are used to compute VIF as a ratio of mutual information measures between the reference and test images. For more details, we refer the reader to \cite{ref:vif}. When using approximation subbands at each level \(\lambda\) to compute VIF, we denote it by \(\text{VIF-A}_\lambda\), and it is computed as:
\begin{equation}
    \text{VIF-A}_\lambda = \frac{\sum\limits_{i,j}\log\left(1 + \frac{g_{\lambda, A}^2(i,j)\sigma_{x, \lambda, A}^2(i,j)}{\sigma^2_{v, \lambda, A}(i,j) + \sigma_n^2}\right)}{\sum\limits_{i,j}\log\left(1 + \frac{\sigma_{x, \lambda, A}^2(i,j)}{\sigma_n^2}\right)},
\end{equation}
where \(\sigma_n^2\) is a small constant used to model neural noise. Note that despite the change in notation, these features are identical to the ``VIF-Scale'' features used in \cite{ref:funque}.

On the other hand, the definition of VIF in \cite{ref:vif} involves applying a vector GSM model on horizontal and vertical subbands of a multi-scale steerable pyramid wavelet decomposition and combining features from all scales. Therefore, analogous to this original formulation of VIF, we compute an \(L\)-level ``scalar-GSM VIF'' on the \(H\) and \(V\) subbands as
\begin{equation}
    \text{VIF-HV}_L = \frac{\sum\limits_{\lambda \leq L}\sum\limits_{\theta \in \{H, V\}}\sum\limits_{i,j}\log\left(1 + \frac{g_{\lambda, \theta}^2(i,j)\sigma_{x, \lambda, \theta}^2(i,j)}{\sigma^2_{v, \lambda, \theta}(i,j) + \sigma_n^2}\right)}{\sum\limits_{\lambda \leq L}\sum\limits_{\theta \in \{H, V\}}\sum\limits_{i,j}\log\left(1 + \frac{\sigma_{x, \lambda, \theta}^2(i,j)}{\sigma_n^2}\right)}.
\end{equation}

A key drawback of VIF is that it is only a measure of spatial quality, and does not include a temporal component. The Spatio-Temporal Reduced Reference Entropic Differencing (ST-RRED) \cite{ref:strred} quality model improves on VIF in this regard by defining measures of both spatial and temporal quality. Inspired by ST-RRED, which is defined in the steerable pyramid \cite{ref:steer} domain on both frames and frame differences, we compute multi-scale ST-RRED features on the approximation subbands of the Haar transform. 

To explain how this is done, we first discuss the computation of the spatial component, termed spatial RRED (SRRED). Using the same local statistics as used by VIF, local information-weighted entropies in subband \(\theta\) at level \(\lambda\) are computed as 
\begin{equation}
   h_{\lambda, \theta}(i, j) = \alpha_{\lambda, \theta}(i,j)\log\left(2\pi e(\sigma^2_{\lambda, \theta}(i, j) + \sigma^2_n)\right),
\end{equation}
where the weighting factors are given by
\begin{equation}
   \alpha_{\lambda, \theta}(i, j) = \log\left(1 + \sigma^2_{\lambda, \theta}(i, j)\right).
\end{equation}

As with our treatment of VIF above, we consider two versions of SRRED. The first is defined on the approximation subbands at various scales, similar to VIF-A.
\begin{equation}
    \text{SRRED-A}_\lambda = \frac{1}{MN}\sum\limits_{i,j}\left| h_{x, \lambda, A}(i, j) - h_{y, \lambda, A}(i, j)\right|.
\end{equation}

The second definition is similar to the definition of VIF-HV, where SRRED is computed on the \(H\) and \(V\) subbands at all levels.
\begin{equation}
    \text{SRRED-HV}_L = \frac{1}{MN}\sum\limits_{\lambda, \theta}\sum\limits_{i,j}\left| h_{x, \lambda, \theta}(i, j) - h_{y, \lambda, \theta}(i, j)\right|.
\end{equation}

To compute Temporal RRED (TRRED), a similar analysis is carried out on the differences between approximation subbands from adjacent frames. Let the local statistics of these ``differenced subbands'' be denoted by \(m_{\lambda, \theta}\), \(s^2_{\lambda, \theta}\), and \(s_{xy, \lambda, \theta}\). Then, the local weighted entropies are given by
\begin{equation}
   g_{\lambda, \theta}(i, j) = \beta_{\lambda, \theta}(i, j) \log\left(2\pi e\left(s^2_{\lambda, \theta}(i,j) + \sigma^2_n\right)\right),
\end{equation}
where the weighting factors are
\begin{equation}
   \beta_{\lambda, \theta}(i, j) = \log\left(1 + \sigma^2_{\lambda, \theta}(i,j)\right) \log\left(1 + s^2_{\lambda, \theta}(i,j)\right).
\end{equation}
The two variants of TRRED are then given by
\begin{equation}
    \text{TRRED-A}_\lambda = \frac{1}{MN}\sum\limits_{i,j}\left| g_{x, \lambda, A}(i, j) - g_{y, \lambda, A}(i, j)\right|.
\end{equation}
and
\begin{equation}
    \text{TRRED-HV}_L = \frac{1}{MN}\sum\limits_{\lambda, \theta}\sum\limits_{i,j}\left| g_{x, \lambda, \theta}(i, j) - g_{y, \lambda, \theta}(i, j)\right|.
\end{equation}

Finally, the combined STRRED features are arrived at:
\begin{equation}
    \text{STRRED-A}_\lambda = \text{SRRED-A}_\lambda \times \text{TRRED-A}_\lambda
\end{equation}
and
\begin{equation}
    \text{STRRED-HV}_L = \text{SRRED-HV}_L \times \text{TRRED-HV}_L.
\end{equation}

\subsection{Detail Loss Metric}
\label{sec:dlm}
The Detail Loss Metric \cite{ref:dlm} is a wavelet-domain full-reference quality model that measures the amount of ``detail loss'' suffered by a test video frame in comparison to a reference video frame. In the following, we summarize the DLM algorithm. Note that due to the use of the shared HVS-sensitive unified transform, we omit the ``contrast sensitivity function'' step described in \cite{ref:dlm}.

The first step in computing DLM involves applying a ``decoupling step.'' The decoupling step is based on the following distortion model used to describe wavelet subband coefficients \(\theta \in \{H, V, D\}\). Let the images \(x\) and \(y\) be reference and test images, respectively. The distortion model assumed by DLM is given by
\begin{equation}
    Y_{\lambda, \theta}(i, j) = \gamma_{\lambda, \theta}(i, j) X_{\lambda, \theta}(i,j) + A_{\lambda, \theta}(i,j),
\end{equation}
where the gain factor \(\gamma\) models attenuation of local gradients due to detail loss, \(R_{\lambda, \theta}(i,j) = \gamma_{\lambda, \theta}(i, j) X_{\lambda, 1}(i,j)\) are the ``restored'' coefficients, and \(A_{\lambda, \theta}(i,j)\) are the ``additive impairments.''

The ``restored'' wavelet decomposition is computed from the given frames as
\begin{equation}
    \hat{R}_{\lambda, \theta}(i,j) = \hat{\gamma}_{\lambda, \theta}(i,j) X_{\lambda, \theta}(i,j),
\end{equation}
where
\begin{equation}
    \psi_{x, \lambda}(i,j) = \arctan\left(\frac{X_{\lambda, V}(i,j)}{X_{\lambda, H}(i,j)}\right),
\end{equation}
\begin{equation}
    \psi_{y, \lambda}(i,j) = \arctan\left(\frac{Y_{\lambda, V}(i,j)}{Y_{\lambda, H}(i,j)}\right),
\end{equation}
\begin{equation}
    \Delta\psi_{\lambda}(i, j) = \left|\psi_{x, \lambda}(i, j) - \psi_{y, \lambda}(i, j) \right|,
\end{equation}
and 
\begin{equation}
    \hat{\gamma}_{\lambda, \theta}(i, j) = \begin{cases}
        \frac{Y_{\lambda, \theta}(i,j)}{X_{\lambda, \theta}(i,j)}, & \Delta\psi_{\lambda}(i, j) < 1^{\circ} \\
        \text{clip}\left(\frac{Y_{\lambda, \theta}(i,j)}{X_{\lambda, \theta}(i,j)}, 0, 1\right), & \textit{else}
    \end{cases}.
\end{equation}

The quantity \(\Delta\psi\) is used to preserve contrast enhancement, which scales both the horizontal and vertical subband coefficients of the distorted frame. The additive impairment coefficients are then computed as
\begin{equation}
    \hat{A}_{\lambda, \theta}(i,j) = Y_{\lambda, \theta}(i,j) - \hat{R}_{\lambda, \theta}(i,j).
\end{equation}

The additive impairments are used to mask the restored coefficients as
\begin{equation}
    \Tilde{R}_{\lambda, \theta}(i,j) = \left(\hat{R}_{\lambda, \theta}(i,j) - M_{\lambda}(i,j)\right)^+,
\end{equation}
where
\begin{equation}
    M_{\lambda}(i,j) = \sum_{\theta}\sum\limits_{i-1}^{i+1}\sum\limits_{j-1}^{j+1} w_{ij}(m,n)\left|\hat{A}_{\lambda, \theta}(m,n)\right|,
\end{equation}
\begin{equation}
    w_{ij}(m,n) = \frac{1 + \delta(m-i, n-j)}{30},
\end{equation}
\((\cdot)^+\) denotes clipping negative values to zero, and \(\delta\) is the Kronecker delta function.

Unlike the formulation used in VMAF and FUNQUE, we compute DLM on a ``per-scale'' basis. The value at scale \(\lambda\) is computed as
\begin{equation}
    \text{DLM-S}_\lambda = \frac{\sum\limits_{\theta \in \{H, V, D\}}\left(\sum\limits_{i,j}\Tilde{R}_{\lambda, \theta}(i,j)^3\right)^{1/3}}{\sum\limits_{\theta \in \{H, V, D\}}\left(\sum\limits_{i,j}\left|X_{\lambda, \theta}(i,j)\right|^3\right)^{1/3}}.
\end{equation}

During feature selection, the ``DLM'' feature bucket allows the selection of either all DLM scales \(\left(\text{DLM-S}_\lambda; \lambda \leq L\right)\), or only the coarsest scale \(\text{DLM-S}_L\). Due to the multi-scale nature of both MS-ESSIM and the information-theoretic features in VIF and ST-RRED, the single-scale DLM offers a good lightweight feature that does not compromise on accuracy.

\subsection{No-Reference Features}
\label{sec:noref}
While full-reference (FR) quality models like the ones discussed so far compute local measures of similarity or distortion between reference and test frames, no-reference (NR) quality models compute measures of ``naturalness'' or quality directly from test frames. Other than recent deep-learning methods \cite{ref:google_nr_dl} \cite{ref:patch_vq}, the two classical approaches to NR quality assessment involve natural scene statistics (NSS) models and artifact measurement.

NSS-based NR quality models typically involve constructing multi-scale multi-orientation band-pass decompositions and building statistical models of the transform coefficients. While such models have proven to be powerful predictors of quality \cite{ref:brisque} \cite{ref:chipqa}, they often face difficulty in generalizing across databases containing highly diverse user-generated content (UGC). 

Artifact-measurement methods such as TLVQM \cite{ref:tlvqm} and the method in \cite{ref:artifact_measurement} attempt to characterize the spatial and temporal properties of video distortions, such as motion, blur, sharpness, blockiness, etc. Based on the success of the TI feature in VMAF and its derivatives, as well as the usefulness of the SI feature in Enh-VMAF, we posit that such features may be useful despite not being based on perceptual/statistical principles.

Keeping our desire for low complexity in mind, we construct two features based on TLVQM - the spatial activity index (TL-SAI) and blurriness (TL-Blur). In TLVQM, both features were computed using Sobel filters \cite{ref:sobel}, but we compute these features using the horizontal and vertical high-pass filters and subbands as gradient functions and responses respectively.

The gradient energy is computed as
\begin{equation}
    E^{(1)}_\lambda = H_\lambda^2 + V_\lambda^2
\end{equation}
and the second-order gradient energy is computed as
\begin{equation}
    E^{(2)}_\lambda = \left(f_{HP} \circledast_H H_\lambda\right)^2 + \left(f_{HP} \circledast_V V_\lambda\right)^2,
\end{equation}
where \(f_{HP}\) denotes the wavelet high-pass filter and \(\circledast_H\) and \(\circledast_V\) denote convolution in the horizontal and vertical directions, respectively.

Using these quantities, the two features are computed as
\begin{equation}
    \text{TL-SAI}_\lambda = \text{std}\left(\sqrt{E^{(1)}_\lambda}\right)^{1/4}
\end{equation}
and
\begin{equation}
    \text{TL-Blur}_\lambda = \frac{\text{mean}\left(\text{perc}\left(E^{(2)}_\lambda, 1\right)\right)}{\text{mean}\left(\text{perc}\left(E^{(1)}_\lambda, 1\right)\right)},
\end{equation}
where \(\text{perc}(\cdot, 1)\) denotes the set of the largest 1\% of values.

NR quality models tend to achieve lower accuracy in comparison to FR quality models since they lack useful information regarding the pristine source content. Since FUNQUE+ operates in the FR regime, we incorporate source information into the artifact features by computing them on both the reference and the test frames and using their difference as features for quality prediction.

\begin{equation}
    \Delta\text{TL-SAI}_\lambda = \text{TL-SAI}_{x, \lambda} - \text{TL-SAI}_{y, \lambda}
\end{equation}
\begin{equation}
    \Delta\text{TL-Blur}_\lambda = \text{TL-Blur}_{x, \lambda} - \text{TL-Blur}_{y, \lambda}
\end{equation}

\subsection{Other Features}
\label{sec:others}
In addition to the aforementioned features, we include low-complexity features based on the differences of subband coefficients. As in FUNQUE, we consider an analog of the ``Motion'' feature of VMAF that we compute as the mean absolute difference (MAD) between approximation subbands of successive frames from the reference video. We denote this feature ``MAD-Ref.'' Similarly, we define ``MAD-Dis,'' which is computed similarly using frames from the distorted video. In addition, we include a simple feature ``MAD,'' which is the difference between the approximation subbands of frames from the reference and distorted videos.

In addition to the MAD features, we include the ``blur'' and ``edge'' features from \cite{ref:evmaf}, which are computed as 

\begin{equation}
    \text{Blur}_\lambda =  \frac{1}{MN}\sum\limits_{i,j}\left(\sum\limits_{\theta \in \{H, V, D\}}\left|X_{\lambda, \theta}(i, j)\right| - \left|Y_{\lambda, \theta}(i, j)\right|\right)^{+}
\end{equation}
and
\begin{equation}
    \text{Edge}_\lambda =  \frac{1}{MN}\sum\limits_{i,j}\left(\sum\limits_{\theta \in \{H, V, D\}}\left|Y_{\lambda, \theta}(i, j)\right| - \left|X_{\lambda, \theta}(i, j)\right|\right)^{+}.
\end{equation}

\section{Experiments}
\label{sec:experiments}

\subsection{Databases Used for Subjective Validation}
\label{sec:databases}

\begin{table}[t]
    \caption{Databases used for Model Selection and Evaluation}
    \label{tab:databases}
    \begin{center}
        \begin{tabular}{|c|c|c|c|c|}
            \hline
            Database &  Size & Codec(s) & Scaling? & Bitdepth\\
            \hline
            BVI-HD & 192 & HM & No & 8 \\
            CC-HD & 108 & HM, AV1, VTM & No & 10\\
            CC-HDDO & 90 & HM, AV1 & Yes & 10\\
            IVP & 100 & Dirac, JM, MPEG-2 & No & 8 \\
            MCL-V & 96 & x264 & Yes & 8\\
            NFLX-P & 70 & x264 & Yes & 8 \\
            SHVC & 64 & HM & No & 8, 10 \\
            VQEG & 72 & MPEG-2, JM & No & 8 \\
            \hline
        \end{tabular}
    \end{center}
\end{table}

To perform model selection and to conduct a thorough evaluation of the FUNQUE+ models, we use the following set of eight databases consisting of full HD videos subjected to scaling and compression distortions.
\begin{enumerate}
    \item BVI High Definition Database (BVI-HD) \cite{ref:bvi_hd} - The BVI-HD database consists of 32 1080p reference videos that have been compressed using the HEVC \cite{ref:hevc} codec at various QP values to yield 192 test videos. The database also consists of a ``texture synthesis'' sub-database, which is not used here due to our focus on streaming artifacts.

    \item BVI Codec Comparison Databases (BVI-CC) \cite{ref:cc_hd} - BVI-CC is a codec-comparison database that evaluated the HM \cite{ref:hm}, AV1 \cite{ref:av1}, and VTM \cite{ref:vtm} codecs under three test conditions. Of the three, we use two sub-databases in this work - CC-HD \cite{ref:cc_hd} and CC-HDDO \cite{ref:cc_hddo}, both of which contain nine 10-bit reference videos.
    
    The CC-HD database consists of 108 test videos generated using the three codecs at the native 1080p resolution. The CC-HDDO database consists of 90 videos that were generated using the HM and AV1 codecs, and it includes scaling artifacts in addition to compression. The various scale factors used for each video were obtained using the Dynamic Optimizer \cite{ref:dyn_optimizer}.

    \item IVP Subjective Quality Video Database (IVP) \cite{ref:ivp} - IVP consists of ten 1920\(\times\)1088 source contents that have been compressed using MPEG2 \cite{ref:mpeg2}, Dirac wavelet \cite{ref:dirac}, and JVET JM H.264 \cite{ref:jm} encoders, yielding 100 test videos.

    \item MCL Video Quality Database (MCL-V) \cite{ref:mcl_v} - MCL-V consists of twelve 1080p source contents distorted using the x264 \cite{ref:x264} encoder at four compression levels each. Scaling is introduced by encoding the videos at 720p in addition to the native 1080p resolution, yielding a total of 96 test videos.

    \item Netflix Public Database (NFLX-P) \cite{ref:vmaf} - NFLX-P consists of nine 1080p source contents distorted using the x264 encoder at ten combinations of bitrate and encoding resolution. However, not all combinations are applied to each video, and the database consists of 70 test sequences in total.

    \item SHVC High Definition Database (SHVC) \cite{ref:shvc} - SHVC contains 1080p videos developed by the Motion Picture Experts Group (MPEG) to evaluate the performance of Scalable HEVC encoding (SHVC). The database consists of 64 test videos after removing videos that overlap with other databases and includes both 8-bit and 10-bit content.

    \item VQEG High Definition Database 3 (VQEG) \cite{ref:vqeghd3} - VQEG consists of nine 1080p source contents distorted using the MPEG-2 \cite{ref:mpeg2} and JM \cite{ref:jm} encoders to yield 72 test videos.
\end{enumerate}

The test databases used to develop and validate the FUNQUE+ models cover a wide range of scenarios, using a variety of encoders, including both 8-bit and 10-bit content, and also modeling the effect of spatial scaling on quality. A summary of the salient features of the eight databases is given in Table \ref{tab:databases}.

\subsection{Evaluation Methodology}
\label{sec:evaluation}
Cross-database generalization is a key property that is required of practical learning-based quality models. While models that have been tuned on individual databases are useful in scenarios where a representative database is available, general-purpose quality models must demonstrate robust cross-database generalization. Indeed, part of the success of VMAF may be attributed to its extensive usage off-the-shelf.

The FUNQUE model was evaluated in \cite{ref:funque} by training on the CC-HDDO database and testing on the other seven databases. CC-HDDO is the largest database that was created using more than one codec, and it includes both compression and scaling distortions. However, that evaluation framework could be interpreted as biased towards the CC-HDDO database, since that is what the feature selection and training were performed using. Therefore, to conduct a more thorough evaluation, we applied a comprehensive cross-database evaluation, where every model was trained on each database and then tested on the other seven, leading to a total of 56 train-test pairs.

We use the Spearman Rank Order Correlation Coefficient (SROCC) as the performance metric and Fisher averaging \cite{ref:ensemble_vmaf, ref:funque} to compute the average over the set of train-test pairs. The Pearson Correlation Coefficient (PCC) is another metric that may be used as the performance objective. Since fusion-based models are trained to predict subjective scores, we observe that these models achieve similar SROCC and PCC values. Moreover, using SROCC allows easy comparison with learning-free (``atomic'') models such as SSIM, without resorting to training and testing logistic mapping functions \cite{ref:essim} across databases.

Therefore, given a set of databases \(D\) and a feature set/model \(F\) to be evaluated, the cross-database SROCC is computed as
\begin{equation}
    \text{CrossDB-SROCC}(F, D) = \text{FMean}\left(\{\text{SROCC}(F, D_i, D_j)\}\right),
\end{equation}
where \(\text{SROCC}(F, D_i, D_j)\) denotes the SROCC achieved by training the model on database \(D_i\) and testing on database \(D_j\), and \(\text{FMean}\) denotes the Fisher average \cite{ref:fisher_z}.

\begin{table}[t]
    \centering
    \caption{Runtime Complexities of Feature Selection Methods}
    \label{tab:feature_selection_comp}
    \begin{tabular}{|c|c|}
        \hline
        Method & Runtime Complexity \\
        \hline
        Exhaustive Feature Search & \(\BigO{2^{NK}}\) \\
        Constrained Exhaustive Feature Search & \(\BigO{N^{K}}\) \\
        Greedy Feature Search & \(\BigO{K^2N^2}\) \\
        \textbf{Constrained Greedy Feature Search} & \(\BigO{K^2N}\) \\
        \hline
    \end{tabular}
\end{table}

\subsection{Constrained Greedy Feature Selection}
\label{sec:cgfs}

\begin{algorithm}
    \caption{Constrained Greedy Feature Selection}\label{alg:cgfs}
    \textbf{Input:} \\
    \hspace*{\algorithmicindent} \(S = \{B_1, \dots, B_K\}\) - Feature ``buckets.''\\
    \hspace*{\algorithmicindent} \(D = \{D_1, \dots, D_T\}\) - Databases for cross-database testing.\\
    \textbf{Output:} \\
    \hspace*{\algorithmicindent} \(F_{\textit{greedy}}\) - The greedy-selected feature set.
    \algrule
    \begin{algorithmic} 
        \STATE \(F_{\textit{greedy}} \gets \phi\)
        \STATE \(S_{\textit{avail}} \gets S\)
        \STATE \(srocc_\textit{best} \gets -1\)
        \WHILE {\(S_\textit{avail} \neq \phi\)}
            \STATE \(F_\textit{best} \gets F_\textit{greedy}\)
            \FOR {\(B \in S_\text{avail}\)}
                \FOR {\(f \in B\)}
                    \STATE \(F_\textit{cand} \gets F_\textit{greedy} \cup \{f\}\)
                    \IF {\(\text{CrossDB-SROCC}\left(F_\textit{cand}, D\right) > srocc_\textit{best}\)}
                        \STATE \(srocc_\textit{best} \gets \text{CrossDB-SROCC}\left(F_\textit{cand}, D\right)\)
                        \STATE \(F_\textit{best} \gets F_\textit{cand}\)
                        \STATE \(B_\textit{chosen} \gets B\)
                    \ENDIF
                \ENDFOR
            \ENDFOR
            \IF {\(F_\textit{greedy} \neq F_\textit{best}\)}
                \STATE \(F_\textit{greedy} \gets F_\textit{best}\)
                \STATE \(S_\textit{avail} \gets S_\textit{avail} \setminus \{B_\textit{chosen}\}\)
            \ELSE
                \STATE \textbf{break}
            \ENDIF
        \ENDWHILE
    \end{algorithmic}
\end{algorithm}

\begin{table*}[t]
    \centering
    \caption{Feature Buckets Used for Feature Selection}
    \label{tab:feat_buckets}
    \begin{tabular}{|P{0.1\linewidth}|P{0.24\linewidth}|P{0.14\linewidth}|P{0.15\linewidth}|P{0.1\linewidth}|}
    \hline
    SSIM & Info & DLM & Sharpness & MAD \\
    \hline
    \(\text{SSIM}_L\), \(\text{ESSIM}_L\), \(\text{MS-SSIM}_L\), \(\text{MS-ESSIM}_L\) &
    \(\text{VIF-HV}_L\), \(\left(\text{VIF-A}_\lambda; \lambda \leq L\right)\), \(\text{STRRED-HV}_L\), \(\left(\text{SRRED-HV}_L, \text{TRRED-HV}_L\right)\), \(\left(\text{STRRED-A}_\lambda; \lambda \leq L\right)\), \(\left(\text{SRRED-A}_\lambda, \text{TRRED-A}_\lambda; \lambda \leq L\right)\) &
    \(\text{DLM-S}_L\), \(\left(\text{DLM-S}_\lambda; \lambda \leq L\right)\) &
    \(\text{Blur}_L\), \(\text{Edge}_L\), \(\left(\text{Blur}_L, \text{Edge}_L\right)\), \(\Delta\text{TL-SAI}_L\), \(\Delta\text{TL-Blur}_L\) &
    \(\text{MAD-Ref}_L\), \(\text{MAD-Dis}_L\), \(\text{MAD}_L\) \\
    \hline
    \end{tabular}
\end{table*}

\begin{table*}[b]
    \centering
    \caption{FUNQUE+ Models}
    \label{tab:funque_plus_models}
    \begin{tabular}{|c|p{0.6\linewidth}|c|c|}
    \hline
    Model & Features & CSF & SAST Used? \\
    \hline
    Y-FUNQUE+ & \(\text{MS-ESSIM}_2 + \text{MAD-Ref}_2 + \text{DLM-S}_2\) & NadenauSW & Yes \\
    \hline
    3C-FUNQUE+ & \(\text{Y-MS-ESSIM}_2 + \text{Y-MAD-Dis}_2 + \text{Y-DLM-S}_2 + \text{Y-SRRED-HV}_2 + \text{Y-TRRED-HV}_2 + \text{Cb-Edge}_2 + \text{Cr-MAD}_2 \) & LiSW & Yes \\
    \hline
    FS-Y-FUNQUE+ & \(\text{MS-ESSIM}_2 + \Delta\text{TL-SAI}_2 + \text{MAD-Dis}_2 + \text{DLM-S}_2 + \text{STRRED-HV}_2\) & NadenauSpat & No \\
    \hline
    FS-3C-FUNQUE+ & \(\text{Y-MS-ESSIM}_3 + \Delta\text{Y-TL-SAI}_3 + \text{Y-DLM-S}_3 + \text{Cb-MAD-Dis}_3 + \text{Cb-SRRED-HV}_3 + \text{Cb-TRRED-HV}_3 + \text{Cb-Edge}_3 + \text{Cr-MAD}_3 + \text{Cr-Blur}_3\) & WatsonSW & No \\
    \hline
    \end{tabular}
\end{table*}

Due to the wide range of design choices available to us, it is required to perform both model selection, i.e., choosing a CSF, number of levels, and the application of SAST, and feature selection for each model. As identified during the development of FUNQUE \cite{ref:funque}, the use of exhaustive feature search (EFS) is not feasible due to the large number of features being considered. This matter was addressed by adopting a Constrained Exhaustive Feature Selection (CEFS) approach.

Under this approach, the set of features under consideration was partitioned into four ``buckets,'' and EFS was used under the constraint that at most one feature may be selected from each bucket. This constraint introduces an inductive bias that identifies robust features of each ``type'' and automatically constrains model size, which improves generalization. For the set of features considered by FUNQUE, this was reported to have reduced the search space by a factor of seven.

However, due to the significantly larger pool of features being considered for FUNQUE+, even CEFS proves to be computationally intractable. So, we instead used a Constrained Greedy Feature Selection (CGFS) algorithm that is significantly more scalable than EFS. A similar greedy feature selection (GFS) method was used in \cite{ref:evmaf}, but it does not include the feature constraints that were found to be crucial in FUNQUE. 

A detailed description of the CGFS algorithm is provided in Algorithm \ref{alg:cgfs}. In short, at each step of CGFS, the feature that achieves the greatest increase in cross-database SROCC is added to the selected feature set, while enforcing the bucket constraints. Considering the case of \(K\) buckets having \(N\) features each, the runtime complexities of the four feature selection algorithms are shown in Table \ref{tab:feature_selection_comp}. Note that CGFS retains the inductive bias introduced by CEFS due to the continued use of feature buckets, while scaling more gracefully on larger candidate feature sets.

We considered five ``feature buckets'' in the feature selection process, grouping features by ``type.'' The five types of features are ``SSIM,'' ``Info,'' which denotes information-theoretic quality models, ``DLM,'' ``Sharpness,'' which includes all sharpness and blur-related features, and ``MAD.'' The feature buckets are described in detail in Table \ref{tab:feat_buckets}. Note that entries in Table \ref{tab:feat_buckets} that are in the form of tuples are groups of features that must be included or excluded collectively.

\subsection{Chroma-aware Modeling}
\label{sec:chroma}
A key weakness of SOTA quality models such as MS-SSIM and VMAF is that they operate only on the luma channel. Improvements to VMAF like ST-VMAF and Ens-VMAF also rely on extracting better features from the luma channel. However, the superior performance of chromatic SSIM \cite{ref:essim} and Enh-VMAF motivates the use of features extracted from the Cb and Cr chroma channels. The tradeoff, however, is that building such ``three-channel'' (3C) quality models requires the processing of all three color channels, which increases the computational complexity.

Therefore, in addition to a luma-only Y-FUNQUE+ model, we have also developed a 3C-FUNQUE+ model that utilizes features extracted from all three color channels - Y, Cb, and Cr. To perform feature selection for the 3C-FUNQUE+ model, we created three copies of each feature bucket described in Section \ref{sec:cgfs}, corresponding to the three channels. This yielded a total of fifteen feature buckets to be used for feature selection.
\section{Results}
\label{sec:results}
\begin{table*}[ht]
    \centering
    \caption{Cross-Database SROCC Achieved by Y-FUNQUE+}
    \label{tab:y_funque_plus_crossdb}
    \begin{tabular}{|c|cccccccc|c|}
    \hline
Database & BVI-HD & CC-HD & CC-HDDO & IVP & MCL-V & NFLX-P & SHVC & VQEG & Average \\
\hline
BVI-HD & - & 0.8356 & 0.8774 & 0.9182 & 0.7486 & 0.9408 & 0.9011 & 0.8912 & 0.8844 \\
CC-HD & 0.7990 & - & 0.8776 & 0.9073 & 0.7942 & 0.9233 & 0.9172 & 0.8473 & 0.8752 \\
CC-HDDO & 0.8038 & 0.8606 & - & 0.9212 & 0.7617 & 0.9205 & 0.9150 & 0.8806 & 0.8777 \\
IVP & 0.7982 & 0.8258 & 0.8745 & - & 0.7837 & 0.9412 & 0.8916 & 0.8861 & 0.8674 \\
MCL-V & 0.7420 & 0.8664 & 0.8559 & 0.8917 & - & 0.8984 & 0.8941 & 0.8038 & 0.8579 \\
NFLX-P & 0.7598 & 0.7363 & 0.7927 & 0.9113 & 0.7614 & - & 0.8412 & 0.8855 & 0.8237 \\
SHVC & 0.7938 & 0.8828 & 0.8879 & 0.9047 & 0.7617 & 0.9334 & - & 0.8884 & 0.8750 \\
VQEG & 0.7909 & 0.7648 & 0.8375 & 0.9126 & 0.7415 & 0.9541 & 0.8742 & - & 0.8586 \\
\hline
Average & 0.7849 & 0.8307 & 0.8603 & 0.9100 & 0.7652 & 0.9327 & 0.8930 & 0.8717 & \textbf{0.8660} \\
\hline
    \end{tabular}
\end{table*}
\begin{table*}[ht]
    \centering
    \caption{Cross-Database SROCC Achieved by 3C-FUNQUE+}
    \label{tab:3c_funque_plus_crossdb}
    \begin{tabular}{|c|cccccccc|c|}
    \hline
Database & BVI-HD & CC-HD & CC-HDDO & IVP & MCL-V & NFLX-P & SHVC & VQEG & Average \\
\hline
BVI-HD & - & 0.8834 & 0.9064 & 0.9129 & 0.7754 & 0.9398 & 0.8833 & 0.8896 & 0.8920 \\
CC-HD & 0.7977 & - & 0.9167 & 0.8838 & 0.8348 & 0.9249 & 0.8834 & 0.8762 & 0.8799 \\
CC-HDDO & 0.7961 & 0.8846 & - & 0.9078 & 0.7676 & 0.9269 & 0.8861 & 0.8819 & 0.8736 \\
IVP & 0.7983 & 0.7901 & 0.8648 & - & 0.7902 & 0.9472 & 0.8458 & 0.8812 & 0.8571 \\
MCL-V & 0.7721 & 0.9216 & 0.9112 & 0.8748 & - & 0.9183 & 0.8672 & 0.8757 & 0.8844 \\
NFLX-P & 0.7759 & 0.8127 & 0.8786 & 0.9064 & 0.7874 & - & 0.8209 & 0.8878 & 0.8455 \\
SHVC & 0.7793 & 0.9331 & 0.9345 & 0.8877 & 0.7928 & 0.9204 & - & 0.8820 & 0.8879 \\
VQEG & 0.7893 & 0.8717 & 0.9371 & 0.8721 & 0.8056 & 0.9317 & 0.8556 & - & 0.8766 \\
\hline
Average & 0.7872 & 0.8795 & 0.9101 & 0.8933 & 0.7943 & 0.9306 & 0.8648 & 0.8822 & \textbf{0.8754} \\
\hline
    \end{tabular}
\end{table*}
\begin{table*}[ht]
    \centering
    \caption{Comparison of Average Test Accuracy (SROCC) of FUNQUE+ Models Against The SOTA Baseline Models. \\ \(^*\) denotes Deep Learning models.}
    \label{tab:avg_sroccs}
    \begin{tabular}{|c|cccccccc|c|}
    \hline
Model & BVI-HD & CC-HD & CC-HDDO & IVP & MCL-V & NFLX & SHVC & VQEG  & Average \\
\hline
SSIM & 0.5986 & 0.7179 & 0.8028 & 0.7145 & 0.3971 & 0.7014 & 0.5624 & 0.7357 & 0.6685 \\
LPIPS\(^*\) & 0.6683 & 0.6663 & 0.7335 & 0.6240 & 0.6319 & 0.8277 & 0.4216 & 0.8212 & 0.6921 \\
PSNR & 0.6125 & 0.6145 & 0.7497 & 0.8169 & 0.4630 & 0.7476 & 0.7542 & 0.7516 & 0.7024 \\
FSIM & 0.6860 & 0.6671 & 0.7183 & 0.6945 & 0.6426 & 0.8566 & 0.5842 & 0.7397 & 0.7082 \\
ST-VMAF & 0.7213 & 0.7748 & 0.8011 & 0.5730 & 0.7048 & 0.8536 & 0.8088 & 0.7657 & 0.7603 \\
MS-SSIM & 0.7849 & 0.6909 & 0.7989 & 0.8900 & 0.6781 & 0.8254 & 0.8567 & 0.6403 & 0.7849 \\
Ens-VMAF & 0.7185 & 0.7601 & 0.8003 & 0.7458 & 0.7881 & 0.9008 & 0.7773 & 0.8166 & 0.7956 \\
DISTS\(^*\) & 0.7686 & 0.6695 & 0.7202 & 0.7930 & 0.7703 & 0.8750 & 0.7979 & 0.8943 & 0.7971 \\
DeepWSD\(^*\) & 0.7979 & 0.7242 & 0.7413 & 0.8756 & 0.7528 & 0.8397 & 0.8304 & 0.8451 & 0.8070 \\
VMAF & 0.7793 & 0.8283 & 0.8633 & 0.8352 & 0.7726 & 0.8924 & 0.8192 & 0.8291 & 0.8312 \\
Enh-VMAF & 0.7644 & 0.7889 & 0.8634 & 0.8703 & 0.7750 & 0.9100 & 0.8383 & 0.8386 & 0.8377 \\
FUNQUE & 0.7624 & 0.7932 & 0.8041 & 0.8895 & 0.7210 & 0.9206 & 0.8608 & 0.8818 & 0.8409 \\
\hline
MS-ESSIM & 0.7693 & 0.7565 & 0.8482 & 0.9026 & 0.6485 & 0.9110 & 0.8874 & 0.8904 & 0.8441 \\
FS-Y-FUNQUE+ & 0.7675 & 0.8220 & 0.8619 & 0.8892 & 0.7159 & 0.9110 & 0.8891 & 0.8499 & 0.8484\\
Y-FUNQUE+ & 0.7849 & 0.8307 & 0.8603 & 0.9100 & 0.7652 & \textbf{0.9327} & \textbf{0.8930} & 0.8717 & 0.8660 \\
\textbf{FS-3C-FUNQUE+} & \textbf{0.7996} & 0.8501 & \textbf{0.9155} & \textbf{0.9125} & 0.7332 & 0.9229 & 0.8602 & \textbf{0.8836} & 0.8707 \\
\textbf{3C-FUNQUE+} & 0.7872 & \textbf{0.8795} & 0.9101 & 0.8933 & \textbf{0.7943} & 0.9306 & 0.8648 & 0.8822 & \textbf{0.8754} \\
\hline
    \end{tabular}
\end{table*}

\subsection{The FUNQUE+ Models}
In this section, we report and analyze the performance of the FUNQUE+ models, and compare them against SOTA baseline models. The luma-only (Y) and three-channel (3C) FUNQUE+ models identified using CGFS are described in Table \ref{tab:funque_plus_models}. In addition, we also describe ``Full-Scale'' models that do not utilize SAST. As a result, these models are less perceptually accurate but are better suited to measure pixel fidelity. 

Furthermore, both Y-FUNQUE+ and 3C-FUNQUE+ utilize subband-weighting CSFs, which are much more computationally efficient than the 21-tap spatial filter used in FUNQUE \cite{ref:funque}. Therefore, overall, both the models are computationally very efficient. This aspect of the FUNQUE+ models is analyzed in further detail in Section \ref{sec:complexity}.

The Full-Scale models are less efficient, both because they are applied at the native 1080p resolution, and since FS-Y-FUNQUE+ uses a spatial filter CSF (albeit only 5-tap). In addition, the two FS models also utilize a larger set of features than the models constructed using SAST.

A closer look at the selected feature sets reveals that all four models utilize MS-ESSIM, which improves upon the ESSIM feature used by FUNQUE. As reported in Table \ref{tab:avg_sroccs}, MS-ESSIM achieves high prediction accuracy even as a standalone quality model. Therefore, the development of MS-ESSIM is one of the main contributions of this work. The high accuracy of MS-ESSIM is attributed to the use of SAST and CoV pooling, similar to ESSIM, as discussed in \cite{ref:essim}. It may also be observed that three of the four models utilize other features drawn from SRRED, TRRED, and \(\Delta\text{TL-SAI}\). In addition, the two 3C models utilize the blur and edge features proposed in \cite{ref:evmaf}. Together, these observations demonstrate the impact of the expanded feature set that was developed to create FUNQUE+.

\subsection{Cross-Database Results}
\label{sec:cross_db}
We tabulate the detailed cross-database testing results of the Y- and 3C-FUNQUE+ models in Tables \ref{tab:y_funque_plus_crossdb} and \ref{tab:3c_funque_plus_crossdb} respectively. Each row refers to one choice of the training database, and each column refers to a choice of test database. Table \ref{tab:avg_sroccs} provides a comparison of the average test SROCC achieved by the FUNQUE+ models and the baseline models against which FUNQUE+ is compared. For brevity, we condensed the cross-database SROCC tables into one row for each model, where each column presents the average test accuracy on that database over all choices of training databases, and the ``Average'' column presents the overall average.

From this table, it may be observed that all of the FUNQUE+ models outperformed all of the baseline models, including VMAF and FUNQUE, as well as more computationally complex models including ST-VMAF and Enh-VMAF. Furthermore, by comparing the performance of the Y-only and 3C models, it may be observed that incorporating the chroma features further improved accuracy. Finally, the FS versions of both models achieved lower SROCC than the other FUNQUE+ models, which demonstrates the tradeoff between modeling viewing conditions and making the model robust to hacking. However, the FS models still significantly outperform the prior SOTA models. 

In addition, one may make two observations regarding the accuracies of baseline models. Firstly, SSIM achieves a lower overall accuracy than all other metrics, even PSNR. While this may appear anomalous, similar observations were made in \cite{ref:evmaf}. Moreover, the higher accuracy of MS-SSIM indicates that the single-scale nature of SSIM is responsible for its lower accuracy. Secondly, all deep methods achieve lower accuracy than VMAF, with LPIPS even dipping below PSNR. We attribute this to the fact that these methods were not designed for the assessment of compression quality. Rather, they often focus on aspects such as geometric and texture distortions.

\subsection{Ablation Analysis}
To understand the performance of FUNQUE+, we conducted ablation experiments by identifying three key design elements that characterize the suite of FUNQUE+ models.
\begin{enumerate}
    \item The incorporation of perceptual sensitivity using a CSF model.
    \item The use of SAST to account for viewing conditions.
    \item The addition of chroma features.
\end{enumerate}
It is important to note that the inclusion of chroma features triples the size of the candidate feature set. Therefore, the development of ``three-channel'' models is contingent upon the use of the scalable CGFS feature selection method described in Section \ref{sec:cgfs}. As a result, the improvement in accuracy by including chroma features also reflects the usefulness of CGFS. To quantify the impact of each of the three factors, we evaluate 8 models that correspond to whether a CSF was used, SAST was used, and whether chroma features were included.

The cross-database accuracy achieved by the eight models is presented in Table \ref{tab:ablation}. From the Table, it may be observed that the addition of each of the three factors improves the accuracy of the quality model in all cases. While including the CSF and chroma features comes at additional computational cost, using SAST improves both accuracy and efficiency, due to downscaling by a factor of 2. In this way, FUNQUE+ attains superior accuracy through the use of CSF, SAST, and chroma features.

\begin{table}[ht]
    \centering
    \caption{Average Test SROCC Achieved by FUNQUE+ Variants}
    \label{tab:ablation}
    \begin{tabular}{|c|c|c|c|}
    \hline
    CSF Used? & SAST Used? & Chroma Included? & Test SROCC \\
    \hline
    No & No & No & 0.8441 \\
    No & No & Yes & 0.8527 \\
    No & Yes & No & 0.8570 \\
    No & Yes & Yes & 0.8640 \\
    Yes & No & No & 0.8484 \\
    Yes & No & Yes & 0.8707 \\
    Yes & Yes & No & 0.8660 \\
    Yes & Yes & Yes & 0.8754 \\
    \hline
    \end{tabular}
\end{table}

\subsection{Computational Optimization and Analysis}
\label{sec:complexity}
The FUNQUE+ models are made quite efficient by sharing computation by calculating all features from a common wavelet decomposition, by restricting model size during the feature selection process, by using lightweight models of the CSF, and by optimizing the implementations of the atom features for efficiency. Note that the optimizations discussed in this section were applied only to the FUNQUE and FUNQUE+ models, not any baseline models that were used for evaluation.

In particular, we used integral images \cite{ref:essim} to remove computationally expensive filtering operations from all features. In this way, all local statistics in the VIF and STRRED algorithms were computed using uniform rectangular windows instead of the usual Gaussian windows.

Filtering using rectangular kernels was eliminated by computing local sums using integral images as follows. Let \(X(i,j)\) denote an image whose local sums over \(k\times k\) windows are to be calculated. First, construct the integral image
\begin{equation}
    I(i,j) = \begin{cases}
                    \sum\limits_{m\leq i}\sum\limits_{n\leq j} X(m,n) & i,j > 0 \\
                    0 & \text{otherwise}
                \end{cases}.
    \label{eq:integral_image}
\end{equation}
The sum over any \(k\times k\) window with top left corner at \((i,j)\) may be computed as
\begin{align}
    S_k(i,j) &= I(i+k-1,j+k-1) + I(i-1,j-1) \nonumber \\
    &- I(i+k-1,j-1) - I(i-1,j+k-1).
    \label{eq:integral_image_end}
\end{align}

The sum of these optimizations, coupled with the use of SAST by Y-FUNQUE+ and 3C-FUNQUE+, provides significant boosts in the efficiency of the FUNQUE+ models as compared to other SOTA quality models. To demonstrate this, we have analyzed all the models in Table \ref{tab:avg_sroccs} to obtain estimates of their computational complexity.

First, we describe the asymptotic computational complexity of the algorithms, expressed in \(\BigO{\cdot}\) (``Big O'') notation in terms of relevant design parameters. In addition, we conducted a thorough analysis of their implementations to estimate the number of Giga Floating Point Operations (GFLOPs) required to run feature extraction for each model on 150 frames of a Full HD video, corresponding to a typical six-second clip played at 25 fps.

\begin{table*}[ht]
    \centering
    \caption{Analyzing the Computational Complexity of FUNQUE+ Against the Baseline Models. \\ GFLOPs Corresponding to Computing Features on 150 Frames of a 1920x1080 Video}
    \label{tab:complexity}
    \begin{tabular}{|c|p{0.6\linewidth}|c|c|}
        \hline
         Model & Asymptotic Computational Complexity & GFLOPs & Time (s) \\
         \hline
         PSNR & \(\BigO{NT}\) & 0.933 & 6.46 \\[5pt]
         \textbf{Y-FUNQUE+} & \(\BigO{k_W\left(1 -2^{-P_W}\right)NT/D^2}\) & 2.245 & 7.68 \\[5pt]
         \textbf{FUNQUE} \cite{ref:funque} & \(\BigO{\left(k_W\left(1 -2^{-P_W}\right) + k_{CSF}^2\right)NT/D^2}\); \(k_{CSF}\): CSF filter size & 16.908 & 13.79 \\[5pt]
         \textbf{3C-FUNQUE+} & \(\BigO{k_W\left(1 -2^{-P_W}\right)NT/D^2}\) & 16.913 & 21.96 \\[5pt]
         SSIM \cite{ref:ssim} & \(\BigO{k_GNT}\) & 75.894 & 29.38 \\[5pt]
         MS-SSIM \cite{ref:ms_ssim} & \(\BigO{k_G(1-2^{-P_G})NT}\) & 104.398 & 43.38 \\[5pt]
         \textbf{FS-Y-FUNQUE+} & \(\BigO{\left(k_W\left(1 -2^{-P_W}\right) + k_{CSF}^2\right)NT}\); \(k_{CSF}\): CSF filter size & 79.393 & 61.37 \\[5pt]
         \textbf{FS-3C-FUNQUE+} & \(\BigO{k_W\left(1 -2^{-P_W}\right)NT}\) & 86.795 & 79.75 \\[5pt]
         VMAF \cite{ref:vmaf} & \(\BigO{\left(k_G\left(1-2^{-P_G}\right) + \left(k_W + k_C^2\right)\left(1 -2^{-P_W}\right)\right)NT}\) & 201.804 & 84.40 \\[5pt]
         Ens-VMAF \cite{ref:ensemble_vmaf} & \(\BigO{\left(\left(k_G + B^4\right)\left(1-2^{-P_G}\right) + \left(k_W + k_C^2\right)\left(1-2^{-P_W}\right)\right)NT}\); \(B\): SpEED block size & 514.018 & 561.52 \\[5pt]
         ST-VMAF \cite{ref:ensemble_vmaf} & \(\BigO{\left(\left(k_G + B^4\right)\left(1-2^{-P_G}\right) + \left(k_W + k_C^2\right)\left(1-2^{-P_W}\right)\right)NT}\); \(B\): SpEED block size & 483.590 & 612.59 \\[5pt]
         Enh-VMAF \cite{ref:evmaf} & \(\BigO{\left(k_G\left(1-2^{-P_G}\right) + \left(k_W + k_C^2\right)\left(1-2^{-P_W}\right) + k_O^2W\left(1 - 2^{-P_O}\right)\right)NT}\); \newline \(k_O\): Optical flow search size, \(P_O\): Optical flow pyramid height, \(W\): Warps per level & 324.896 & 737.77 \\[13pt]
         FSIM \cite{ref:fsim} & \(\BigO{\left(\log(N) + P_{LG}O + k_{Grad}^2\right)NT}\); \(P_{LG}\): Height of log-Gabor pyramid, \newline \(O\): Number of orientations, \(k_{Grad}\): Gradient filter size & 429.484 & 1047.35 \\[13pt]
         LPIPS \cite{ref:lpips} & \(\BigO{\left(\sum\limits_{i=1}^{5}k_i^2C_iC_{i-1/s_i^2}\right)NT}\); \(k_i\): Convolution kernel sizes, \(C_i\): Convolution channels, \newline \(s_i\): Sub-sampling factors due to striding and max-pooling. & 18205.206 & 389.56 \\[13pt]
         DISTS \cite{ref:dists} & \(\BigO{\left(\sum\limits_{i=1}^{5}L_ik_i^2C_i^2/s_i^2 + \sum\limits_{i=1}^{4} k_H^2 C_i/s_i^2\right)NT}\); \(L_i\): Layers in block \(i\), \(k_i\): Convolution kernel sizes, \(C_i\): Convolution channels, \(s_i\): Sub-sampling factors due to striding and max-pooling, \(k_H\): Hanning window size. & 382213.417 & 484.48 \\[13pt]
         DeepWSD \cite{ref:deep_wsd} & \(\BigO{\left(k_R^2 + \sum\limits_{i=1}^{5}L_ik_i^2C_i^2/s_i^2 + \sum\limits_{i=1}^{4} \log\left(W^2\right) C_i/s_i^2\right)NT/s_R^2}\); \(k_R\): Resize kernel size, \(s_R\): Resize factor (8), \(L_i\): Layers in block \(i\), \(k_i\): Convolution kernel sizes, \(C_i\): Convolution channels, \(s_i\): Sub-sampling factors due to striding and max-pooling, \(k_H\): Hanning window size, \(W\): Wasserstein distance block size (16). & 12014.977 & 1589.46 \\[13pt]
         \hline
    \end{tabular}

    \vspace{3pt}
    * Note: Wherever applicable, \(N\) denotes the number of pixels per frame, \(T\) denotes the number of frames in the video, \\
    \(D\) denotes the SAST downscaling factor, \(k_G\), \(k_W\) and \(k_C\) denote the size of Gaussian, wavelet, and contrast-masking filters, \\
    and \(P_G\) and \(P_W\) denote the heights of the Gaussian and wavelet pyramids.
\end{table*}

In addition, we measured the run time of all the compared models on an Intel Core i7-8700 CPU having a clock frequency of 3.2GHz. All of the models were implemented in Python for a fair comparison. The results of the computational analysis are provided in Table \ref{tab:complexity} and the tradeoff between run time and model accuracy has been visualized in Fig. \ref{fig:srocc_v_time}. From these results, it may be seen that the proposed FUNQUE+ (and FUNQUE) models offer the highest average test SROCC, while also being the most efficient fusion-based models. Note that the run time on the horizontal axis is shown in a log scale to accommodate the wide range of run times observed during the experiment.

\begin{figure}
    \centering
    \includegraphics[width=0.9\linewidth]{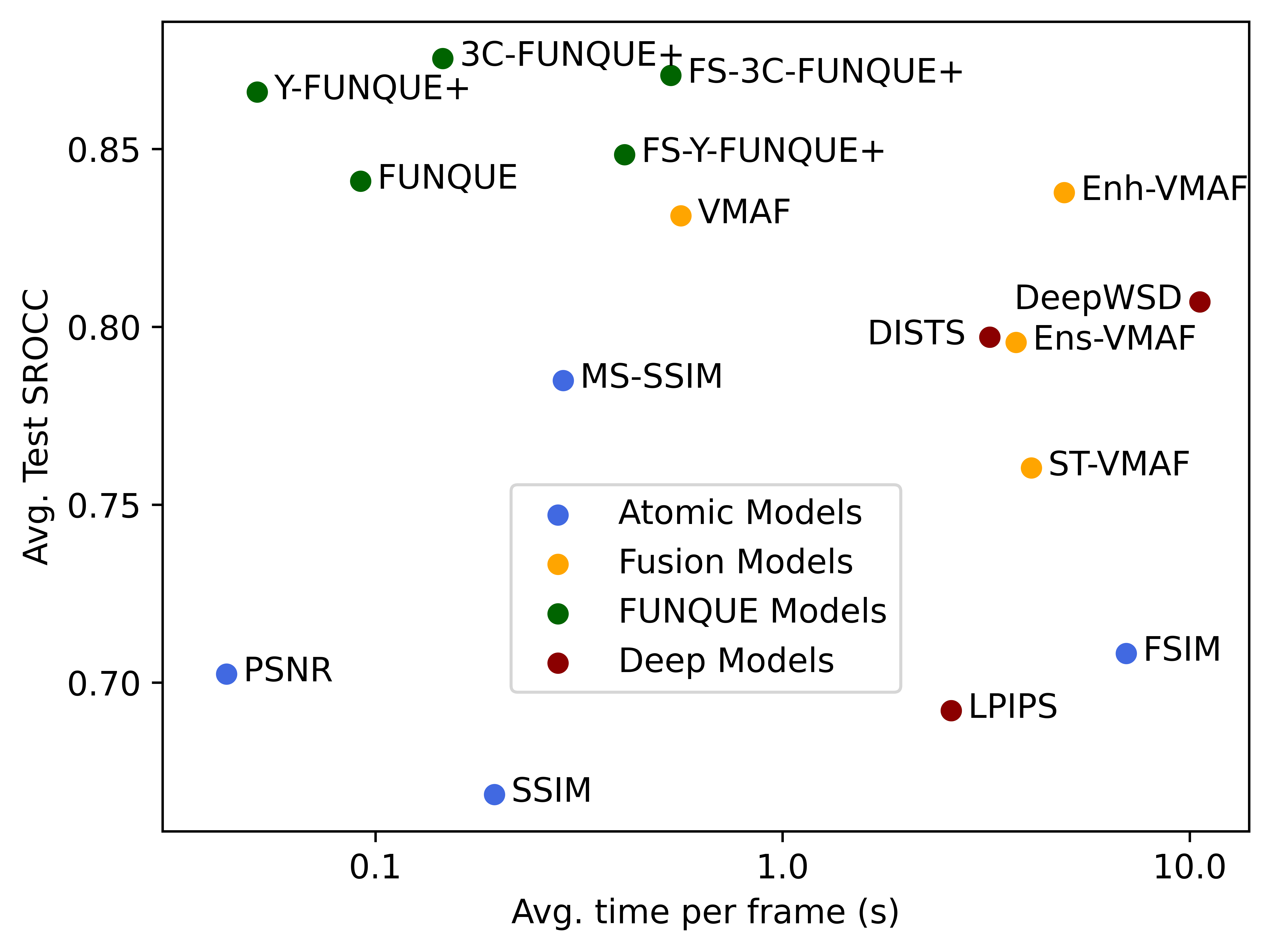}
    \caption{SROCC vs. Run Time}
    \label{fig:srocc_v_time}
\end{figure}

\subsection{Monotonicity Analysis}

\begin{figure*}[ht]
\centering
    \subfloat[VMAF v0.6.1\label{fig:vmaf_monotonicity}]{%
      \includegraphics[width=0.3\linewidth]{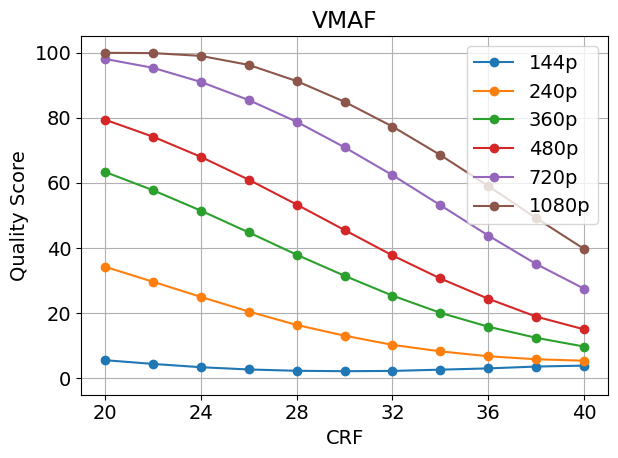}
    }%
    \subfloat[Y-FUNQUE+\label{fig:y_funque_plus_monotonicity}]{%
      \includegraphics[width=0.3\linewidth]{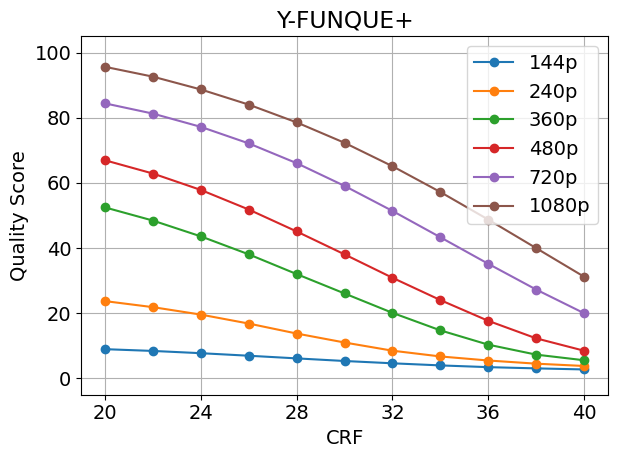}
    }%
    \subfloat[3C-FUNQUE+\label{fig:3c_funque_plus_monotonicity}]{%
      \includegraphics[width=0.3\linewidth]{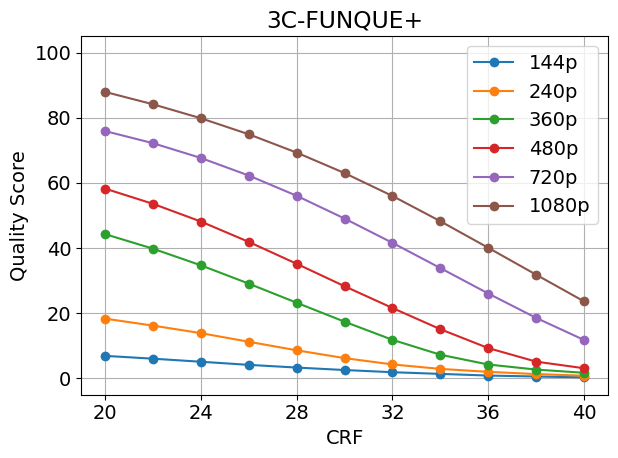}
    }
    \caption{Quality Predictions Made by VMAF, Y-FUNQUE+, and 3C-FUNQUE+ on the ``Seeking'' Source Content.}
    \label{fig:seeking_preds}
\end{figure*}

Finally, we evaluate the monotonic behavior of the proposed FUNQUE+ models. To be suitable for use in perceptual rate-distortion optimization, a quality model must satisfy two basic monotonicity properties related to compression. Let \(Q(r, c)\) denote the prediction of a video quality model for a video compressed at resolution \(r\) and a given compression factor \(c\) (e.g., CRF in libx264). Then, the model is monotonic if whenever \(r_1 \leq r_2\), then \(Q(r_1, c) \leq Q(r_2, c)\), and whenever \(c_1 \leq c_2\), where lower \(c\) denotes less compression, then \(Q(r, c_1) \geq Q(r, c_2)\).

To analyze the monotonicity of the FUNQUE+ models, we used the nine 8-bit 1080p source contents from the NFLX-P database. Each source video was encoded at six encoding resolutions - 1080p, 720p, 480p, 360p, 240p, and 144p and eleven CRF values equally spaced in the range 20-40 using the x264 encoder, yielding a total of 792 test videos. Quality predictions were made using VMAF v0.6.1 and the two FUNQUE+ models Y-FUNQUE+ and 3C-FUNQUE+, which were trained on the CC-HDDO database. The CC-HDDO database was chosen for this analysis for the reasons described in Section \ref{sec:evaluation}, and also because it does not include the x264 encoder. 

The monotonicity of the predictions made by all three models was analyzed both in terms of encoding resolution and CRF. It was observed that for three of the nine source contents, the quality predicted by VMAF v0.6.1 did not decrease with an increase in CRF at low encoding resolutions. An example of this behavior is illustrated in Figure \ref{fig:vmaf_monotonicity}, which shows the variation of predicted visual quality against encoding resolution and CRF for the source video named ``Seeking.'' Interestingly, retraining the VMAF regressor on the CC-HDDO database restored monotonicity to VMAF's predictions. Therefore, we attribute the source of non-monotonicity to the VMAF v0.6.1 regressor and training dataset.

On the other hand, as shown in Figures \ref{fig:y_funque_plus_monotonicity} and \ref{fig:3c_funque_plus_monotonicity}, it was found that both FUNQUE+ models predicted perfectly monotonic variations of quality with CRF at all resolutions. This demonstrates that the FUNQUE+ models are robust even in the low quality regime, despite their efficiency.

\section{Conclusion and Future Work}
\label{sec:conclusion}
We have described a class of efficient fusion-based full-reference video quality models that have been designed to have exceptional practical performance attributes when applied to large-scale video streaming. These models were based on the prototype FUNQUE framework and were developed using novel wavelet-domain features and a novel feature selection algorithm that yields small, diverse fusion-based models. Through extensive analysis and experimental validation, we have been able to show that the resulting FUNQUE+ models achieve higher accuracy than SOTA FR models with significantly reduced computational cost.

We hope that in addition to finding use in the video streaming and sharing industries, the success of the FUNQUE+ models will lead to the development of other low-complexity, high-accuracy models using similar principles. As described in Section \ref{sec:background}, VMAF has grown to find use beyond streaming, particularly in emerging video modalities. In the future, we envision FUNQUE-based models being used across domains, to perceptually optimize the delivery of 360 VR videos, high-resolution videos including 8K, and High Dynamic Range (HDR) videos.

\section{Acknowledgment}
\label{sec:acknowledgment}
The authors would like to thank the Texas Advanced Computing Center (TACC) for supporting this research by providing high-performance computational resources.

\bibliographystyle{IEEEtran}
\bibliography{refs}
\end{document}